%% file: Linden2022_arxiv.tex
\UseRawInputEncoding
\documentclass[twocolumn]{aastex62}

\usepackage{xspace}

\newcommand{\starcnet}{\textsc{StarcNet}\xspace}

\shorttitle{Star Cluster Formation and Evolution in M101: A LEGUS Perspective}
\shortauthors{S. Linden et al.}

\begin{document}

\title{Star Cluster Formation and Evolution in M101: An Investigation with the Legacy Extragalactic UV Survey}

\author{S. T. Linden}
\affiliation{Department of Astronomy, University of Massachusetts at Amherst, Amherst, MA 01003, USA.}

\author{G. Perez}
\affiliation{Department of Astronomy, University of Massachusetts at Amherst, Amherst, MA 01003, USA.}

\author{D. Calzetti}
\affiliation{Department of Astronomy, University of Massachusetts at Amherst, Amherst, MA 01003, USA.}

\author{S. Maji}
\affiliation{Department of Astronomy, University of Massachusetts at Amherst, Amherst, MA 01003, USA.}

\author{M. Messa}
\affiliation{Observatoire de Geneccve, Université de Genève, Chemin Pegasi, 51 1290 Versoix, Switzerland}
\affiliation{Department of Astronomy, Oskar Klein Centre, Stockholm University, AlbaNova University Centre, 106 91 Stockholm, Sweden}

\author{B. C. Whitmore}
\affiliation{Space Telescope Science Institute, 3700 San Martin Drive, Baltimore, MD, USA}

\author{R. Chandar}
\affiliation{University of Toledo, 2801 W. Bancroft St., Mail Stop 111, Toledo, OH, 43606}

\author{A. Adamo}
\affiliation{Department of Astronomy, Oskar Klein Centre, Stockholm University, AlbaNova University Centre, 106 91 Stockholm, Sweden}

\author{K. Grasha}
\affiliation{Research School of Astronomy and Astrophysics, Australian National University, Canberra, ACT 2611, Australia}
\affiliation{ARC Centre of Excellence for All Sky Astrophysics in 3 Dimensions (ASTRO 3D), Australia}

\author{D. O. Cook}
\affiliation{Caltech/IPAC, 1200 E. California Boulevard, Pasadena, CA 91125, USA}

\author{B. G. Elmegreen}
\affiliation{IBM Research Division, T.J. Watson Research Center, Yorktown Heights, NY 10598, USA}

\author{D. A. Dale}
\affiliation{Department of Physics and Astronomy, University of Wyoming, Laramie, WY 82071, USA}

\author{E. Sacchi}
\affiliation{Space Telescope Science Institute, 3700 San Martin Drive, Baltimore, MD, USA}
\affiliation{Leibniz-Institut fur Astrophysik Potsdam, An der Sternwarte 16, 14482 Potsdam, Germany}
\affiliation{INAF Osservatorio di Astrofisica e Scienza dello Spazio di Bologna, Via Gobetti 93/3, I-40129 Bologna, Italy}

\author{E. Sabbi}
\affiliation{Space Telescope Science Institute, 3700 San Martin Drive, Baltimore, MD, USA}

\author{E. K. Grebel}
\affiliation{Astronomisches Rechen-Institut, Zentrum f\"ur Astronomie der Universit\"at Heidelberg, M\"onchhofstr.\ 12--14, 69120 Heidelberg, Germany}

\author{L. Smith}
\affiliation{European Space Agency (ESA), ESA Office, Space Telescope Science Institute, 3700 San Martin Drive, Baltimore, MD 21218, USA}

\date{\today}

\begin{abstract}

We present {\it Hubble Space Telescope} WFC3/UVIS (F275W, F336W) and ACS/WFC optical (F435W, F555W, and F814W) observations of the nearby grand-design spiral galaxy M101 as part of the Legacy Extragalactic UV Survey (LEGUS). Compact sources detected in at least four bands were classified by both human experts and the convolutional neural network \starcnet. Human experts classified the 2,351 brightest sources, retrieving $N_{c} = 965$ star clusters. \starcnet, trained on LEGUS data not including M101, classified all 4,725 sources detected in four bands, retrieving $N_{c} = 2,270$ star clusters. The combined catalog represents the most complete census to date of compact star clusters in M101. We find that for the 2,351 sources with both a visual- and ML-classification \starcnet is able to reproduce the human classifications at high levels of accuracy ($\sim 80-90\%$), which is equivalent to the level of agreement between human classifiers in LEGUS. The derived cluster age distribution implies a disruption rate of $dN/d\tau \propto \tau^{-0.45 \pm 0.14}$ over $10^{7} < \tau < 10^{8.5}$yr for cluster masses $\geq 10^{3.55} M_{\odot}$ for the central region of M101 and $dN/d\tau \propto \tau^{-0.02 \pm 0.15}$ for cluster masses $\geq 10^{3.38} M_{\odot}$ in the northwest region of the galaxy. The trends we recover are weaker than those of other nearby spirals (e.g. M51) and starbursts, consistent with the M101 environment having a lower-density interstellar medium, and providing evidence in favor of environmentally-dependent cluster disruption in the central, southeast, and northwest regions of M101. 

\end{abstract}


\section{Introduction}

The large-scale environments of galaxies play an important role in their evolution: galaxies in denser environments tend to be more bulge dominated, less gas-rich, and have lower overall star formation rates \citep[SFR;][]{tempel11}. These trends are especially pronounced in galaxy clusters where gas can be efficiently removed through tidal interactions and ram pressure stripping \citep[e.g.,][]{hester06,roediger09,boselli21}. However, some of these effects also appear in less dense environments, such as galaxy groups and mergers. To better understand this pre-processing of galaxies, we need a more complete picture of the gas removal and accretion processes across all types of environments.

M101 (NGC 5457) provides a unique laboratory for studying the evolution of a massive spiral galaxy undergoing tidal interactions with two late-type dwarf galaxies (NGC 5474, NGC 5477) within a nearby galaxy group \citep{geller83}.~In such environments, minor companions can drive significant evolution in disk structure and ongoing star formation activity. Observations of M101 and other nearby spiral galaxies indicate that their stellar populations and emission properties can be strongly dependent on radius \citep{walker64, pilyugin14, croxall16, esteban20}. These trends are attributed both to chemical abundance gradients \citep[e.g.,][]{hartwick70, humphreys84, eggenberger02, grammer13}, as well as radial variations in their star formation histories (SFH) \citep[e.g.,][]{smith18, williams18, dale20}. Many of these studies point to a scenario of galaxy evolution where disks grow primarily through star formation triggered by the inflow of pristine intergalactic gas, major mergers, or the accretion of smaller satellite galaxies \citep{governato07}. Observational markers of such an inside-out formation scenario include negative metal abundance gradients and ultraviolet/optical colors that are increasingly blue as a function of galactocentric radius \citep[e.g.,][]{larson76, ryder94, barnes14, dsouza14}.

In addition to radial variations in their stellar populations, nearby spiral galaxies contain a large number young massive stellar clusters (YSCs), which has led to the suggestion that these objects could represent the present-day analogs of ancient globular clusters \citep[e.g.,][]{zepf99,bcm02a}. The conditions under which these clusters can survive for $\geq 10$ Gyr is still uncertain due to a lack of consensus on the influence different galactic environments have on YSC evolution. While cluster destruction through gas removal ($1-10$ Myr) appears largely mass-independent \citep{pfeffer17}, there is not agreement on whether later phases of cluster destruction are mass- or environmentally-dependent. In the case of mass-dependent-disruption \citep[MDD; e.g.,][]{lamers05,nb12} a cluster's lifetime depends on its mass such that low-mass clusters in weaker tidal fields or ones with fewer interactions with surrounding giant molecular clouds (GMCs) have longer lifetimes than low-mass clusters in stronger fields. Whereas, mass-independent disruption \citep[MID; e.g.,][]{bcm07,mf09} suggests that there is a universal expression for the fraction of clusters that are disrupted per unit (logarithmic) time, with roughly $90\%$ of clusters disrupting in each decade of time independent of the cluster mass.

These two scenarios are likely not mutually exclusive: MID may arise in extreme star-forming conditions or with extreme ambient tidal fields like what is seen in the Antennae Galaxies \citep[e.g.,][]{mf05,bcm10}. However, in this case the overall properties of a galaxy's interstellar medium (ISM) may still have a strong influence on the resulting cluster age and mass distributions \citep{bge10}. Several studies have found evidence for environmentally-dependent cluster formation and evolution \citep{boutloukos03,esv11}. \citet{esv14} and \citet{aa17} found a higher disruption rate of clusters toward the center and little or no disruption in the outer regions of M83 and NGC 628 respectively. \citet{mm18b} find a similar trend in M51a (i.e., NGC 5194), which is undergoing an interaction with its lower-mass companion M51b (i.e., NGC 5195). Finally, \citet{linden21} find similar evidence for environmental influence on the disruption rates seen towards the centers of luminous infrared galaxies involved in major mergers in the local Universe. 

One of the primary challenges for discriminating between different disruption scenarios is the lack of large and homogeneously selected samples of star clusters with masses below $\sim 10^{4} M_{\odot}$ in galaxies which contain a large diversity in their star-forming environments. With one of the largest observed negative abundance gradients for a spiral galaxy in the local Universe \citep[$\sim \Delta0.6$dex in lo.pdfg$(O/H)$ over $R_{25}$ - ][]{croxall16}, and a diameter nearly twice that of the Milky Way (MW), the cluster populations in M101 can shed light on the role environment plays on the evolution of star clusters in galaxies. Large numbers of compact clusters have been seen previously in lower-resolution wide-field images with WFPC2 \citep{bresolin96,barmby06} and ACS \citep{simanton15}. However a complete multi-band census of M101's cluster population has remained unexplored relative to other nearby massive spiral galaxies \citep[e.g.,][for a review]{aa20b}. Targeted spectroscopic observations of the most massive clusters and giant HII regions in M101 \citep{chen05, coogan17} reveal that M101 contains both young and old clusters with properties (size, luminosity) similar to clusters in other nearby spiral galaxies and the MW.

Here we present the results from an HST ACS and WFC3 investigation into the star cluster populations of M101 with the Legacy Extragalactic UV Survey \citep[LEGUS:][]{calzetti15}, which provides a two orders of magnitude larger sample of well-characterized star clusters than what has been done thus far. We adopt a distance of 6.7 Mpc for M101 \citep{sabbi18}. The paper is organized as follows: In \S 2, the observations, cluster identification, classification via both humans and machine learning (ML) techniques, and spectral energy distribution (SED) fitting are presented. In \S 3 we compare the human-classified and ML-classified cluster samples and discuss the resulting completeness limits of the cluster sample. In \S 4, the age and mass functions are discussed within the context of nearby massive spiral galaxies for clusters found in both the inner and outer-disk of M101. \S 5 is a summary of the results.

\begin{figure*}
  \centering
  \includegraphics[scale=0.58]{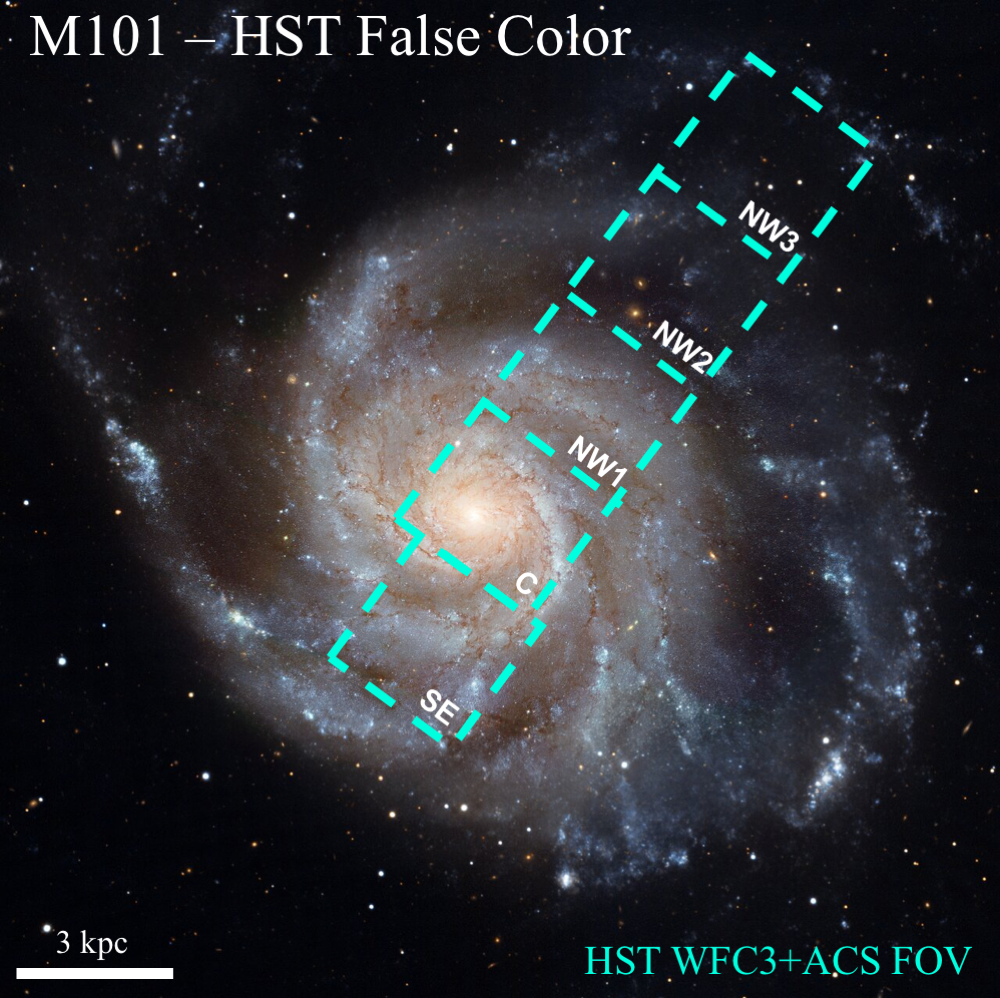}
  \caption{A Hubble Space Telescope (HST) false-color image (courtesy of ESA/Hubble) of M101 is shown overlaid with the combined HST WFC3+ACS FOV composed of 5 pointings extending from the SE to the NW of the galaxy's disk. Our coverage allows us to probe distinctly different star-forming environments between the central regions and the outer-disk. In total 39,705 sources are detected across these 5 fields which satisfy this cut with detections in at least 2 broadband filters which are ultimately extracted and included in our initial photometric catalog.}
\end{figure*}

\section{Observations, Cluster Identification, and Classification}

M101 was observed as part of LEGUS in two UV bands (NUV=F275W, U=F336W) with the UVIS channel on the Wide Field Camera 3 (WFC3) for five independent fields which we label the C, SE, NW1, NW2, and NW3 (see Figure 1). These observations were complemented by archival Wide Field Camera images from the Advanced Camera for Surveys (ACS) in three optical filters (B=F435W, V=F555W, I=F814w) covering the same field-of-view. The pixel scale of the final images in each field is 0.04$"$. The completeness of this photometric dataset allows us to break the age-reddening degeneracy in star clusters, and enable derivation of ages, masses, and extinctions with $\sim 0.1-0.2$ dex accuracy, via SED-fitting on a cluster-by-cluster basis \citep[e.g.,][]{aa17,mm18a,cook19}. Star clusters with masses below $\sim 10^{3} M_{\odot}$ are subject to significant stochastic sampling of the underlying initial mass function (IMF), which primarily affects the ionizing photon rate and thus luminosity of clusters with ages $\leq 10$ Myr \citep[][]{km15,ashworth17}. We will address the mass completeness limit of our derived cluster samples in \S 3.1. A description of the cluster selection, photometry, and SED fitting procedures are further presented in \citet{aa17,grasha19}. We briefly summarize here the key aspects of these methods.

The automated catalogue of star cluster candidates is extracted from the ACS F555W image (PSF $\sim 0.1"$) for each pointing with source extractor \citep[SExtractor;][]{sex}. The SExtractor input parameters are optimized to identify sources with a $\geq 10 \sigma$ detection in 10 continuous pixels. This procedure returns the positions of candidate clusters within each pointing and the concentration index (CI: defined here as the magnitude difference of the source within apertures of 1 and 3 pixels). The CI is related to the effective radius of each object \citep{ryon17} and can be used to differentiate between individual stars and clusters; stars have narrower light profiles and, therefore, smaller CI values compared to star clusters. The CI reference value used to distinguish between unresolved sources (stars) and resolved sources (candidate clusters) within M101 is 1.4 mag \citep{aa17}. There are 39,705 sources which satisfy this cut and are ultimately extracted and included in our initial photometric catalog.

Standard aperture photometry is performed for each cluster candidate using a fixed 4 pixel radius with a local sky annulus from 7-8 pixels in all five filters. Aperture corrections to account for missing flux are based on isolated clusters in each field separately \citep[see][]{mm18a} and calculated by subtracting the photometry in our fixed apertures from the total photometry inside a 20 pixel aperture with a 1 pixel-wide sky annulus. Finally, corrections for foreground Galactic extinction \citep{schlafly11} are applied to the photometric measurements. Following this, all cluster candidates detected in 4-5 bands, which also have low U-band photometric uncertainties ($\sigma_{U} \leq 0.1$ mag) undergo SED fitting to derive the age, mass, and color excess E (B-V) of each source. This reduces the final sample of \starcnet classified sources to 4,725 and ensures all objects have at least one detection in a NUV- or U-band filter; this is critical for achieving the accuracies stated above in the derived cluster age, mass, and extinction.

The SED fitting is performed using the Yggdrasil single stellar population (SSP) models, which implement Cloudy to include the contribution from nebular emission lines \citep{yggdrasil}. All cluster catalogues for the LEGUS galaxies use a \citet{pk01} initial mass function (IMF), and the cluster properties in this paper are derived using Padova isochrones that include thermally pulsating asymptotic giant branch stars \citep{vazquez05} as well as a MW attenuation curve. We have additionally performed all of the analysis described in \S 4.1 using the Geneva isochrones \citep{geneva12} which are often better-suited for modeling populations of very young clusters found in starburst galaxies. We find  minimal impact on the derived quantities from the change in isochrones, with changes in those quantities that are comparable or below our reported uncertainties. Our least-$\chi^{2}$ SED fitting produces average uncertainties of 0.2 dex in both the cluster ages and masses as long as the number of filters used is 4 or more \citep{aa17}. Since our cluster photometry is measured by applying a single average aperture correction in each filter, its uncertainty affects only the normalization of the SED, and not the overall shape. As a result, our average aperture corrections only introduce a 0.2 dex uncertainty in the derived cluster mass \citep{doc19}.

\begin{figure*}
  \centering
  \includegraphics[scale=0.28]{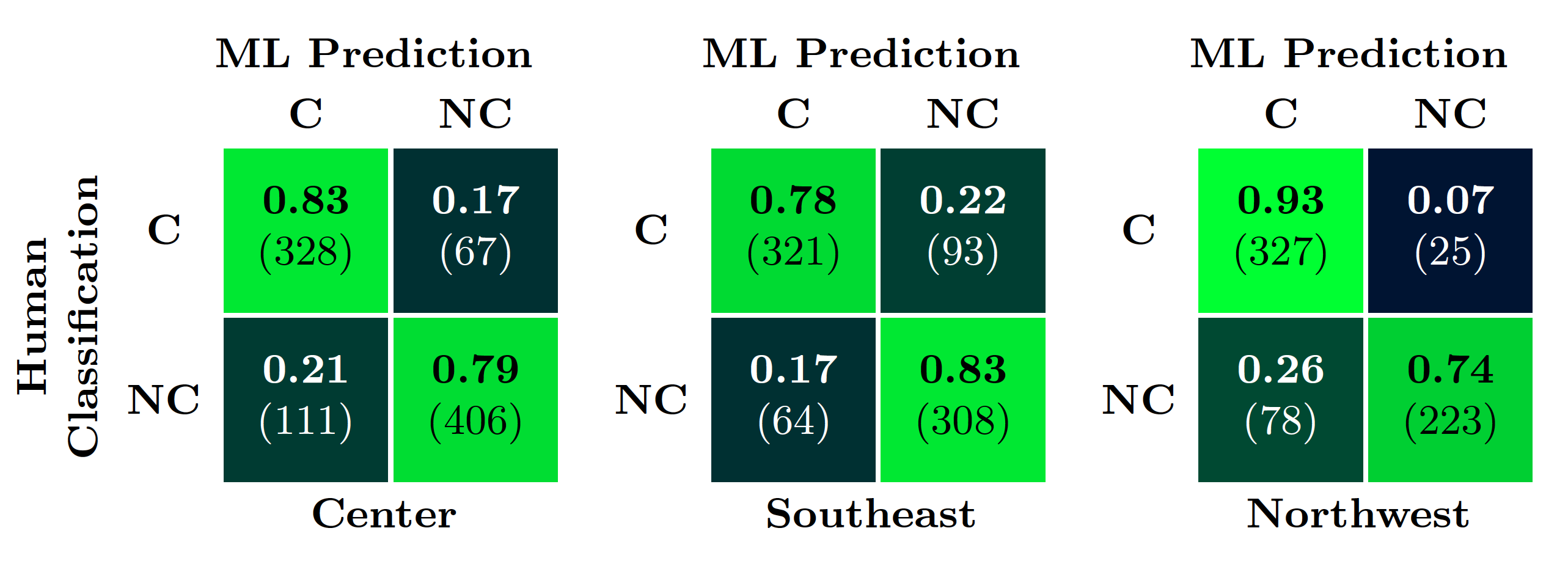}
  \caption{Confusion Matrix for our visually classified clusters vs ML predictions from \starcnet binned as Class 1 and 2, which we adopt as our cluster sample (C), and Class 3 and 4, which we assign as non-clusters (NC). The ability for \starcnet to recover Class 1 and 2 objects in M101 at the same levels or better than human classified catalogs \citep{perez21} is clearly demonstrated, and also suggests the model performs slightly better in less crowded regions of the galaxy.}
\end{figure*}

\subsection{Visual Classification}

In LEGUS, cluster candidates were placed into four classes based on their morphological appearance \citep[see][]{grasha15,aa17,doc19,bcw20}. Class 1 objects are compact, symmetric, single peaked, and extended relative to stellar point sources in the image. Class 2 objects are asymmetric but otherwise share the same properties as Class 1 objects. Class 3 objects contain multiple peaks with the same color, and are often associated with binary or multiple stellar sources or compact OB associations. We also note that in some cases Class 3 objects contain diffuse (nebular) emission associated with nearby star forming regions. Class 4 objects are not star clusters nor associations, and generally consist of background galaxies, individual stars, chance groupings, or image artifacts (cosmic rays, bad pixels, etc.). In order to obtain a reasonable set of clusters for humans to visually inspect a magnitude threshold of $m_{V} = 23.14$ ($M_{V} = -5.99$) was additionally applied to the catalog. In total 2,351 objects were visually classified by two LEGUS team members independently (Sean Linden and Bradley Whitmore) to obtain a consensus result. For the analysis in this study we are interested only in Class 1 and 2 sources which we treat as a single type of object, a likely-bound star cluster well described by SSP models. OB associations and larger star-forming regions which may be un-bound and contain multiple generations of star formation are generally given Class 3 designations. Roughly $\sim 50\%$ of the objects visually inspected were determined to be Class 1 or 2 clusters, which is consistent with the fractional number of Class 1 and 2 objects found for the majority of the LEGUS cluster catalogs \citep[see Figure 2 of][]{grasha17}, and therefore represents the final catalog of human-generated star clusters in M101.

\subsection{Machine Learning Classification}

\starcnet is a three-pathway convolutional neural network (CNN) developed to classify star clusters in LEGUS five-band images \citep{perez21}. The model was trained using the combined dataset of all human-classified cluster catalogs publicly released as part of LEGUS. Crucially, this dataset does not include M101, and thus the machine learning classifications for M101 in this work are independent of the pre-trained \starcnet model. Over 15,000 sources across 31 galaxies, visually classified by experts from the LEGUS team, were used to train and test \starcnet. The final version of the model reaches an overall accuracy of $\sim 69\%$, which approximately matches the agreement achieved among the experts who examined these galaxies ($\sim 70 - 75\%$). However this accuracy is not uniform across classes, with a better performance being achieved for Class 1 and 4 objects; highlighting the difficulty for human classifiers to accurately identify objects of Class 2 and 3. The accuracy of the model increases to $86\%$ when re-measured to adopt the binary classification of clusters (Class 1 and 2) vs. non-clusters (Class 3 and 4) that we are using in this study \citep{perez21}. 

Using this model we generate a classification for all 4,725 sources with detections in 4 or more filters and low U-band photometric uncertainties across our 5 WFC3 and ACS fields in M101. We stress that, opposite to the LEGUS visual classification, we do not apply a magnitude threshold to the sources that are classified by \starcnet. For the remainder of our analysis we separate these fields into three different regions which correspond to the central (C), southeast (SE), and three northwest pointings (NW) in M101. In Figure 2 we compare the confusion matrix for the 2,351 clusters which have both a visual and machine learning (ML) generated classification. Ultimately we have a relatively uniform distribution of Class 1 and 2 clusters across our three catalogs, and find that the agreement between the two methods is high ($78-92\%$). With M101 excluded from the original training data used to develop \starcnet, this level of agreement demonstrates that \starcnet can produce human-level classification accuracy for any sample of objects detected in a galaxy out to $\sim 10$ Mpc for which the broadband filters used as part of LEGUS are available. Further, our results for M101 are consistent with comparisons of both human- and machine-learning generated cluster catalogs produced as part of the Physics at High Angular Resolution Nearby Galaxy Survey \citep[PHANGS;][]{bcw21}. We note that similar to previous star clusters catalogs produced using machine-learning algorithms \starcnet does not identify many faint Class 3 objects. From visual inspection we find that these objects are likely being redistributed into all of the other classes fairly equally, and therefore do not bias the resulting age and mass distributions. 

There also appears to be a higher level of agreement for the NW catalog which probes a region of M101 which is much less crowded, and therefore likely to make it easier to identify and classify individual objects. Upon further inspection of objects in the lower-left and upper-right cross-diagonal of Figure 2 (i.e. objects \starcnet determines to be clusters that were visually inspected and placed in Class 3 or 4) we find that many of them are faint Class 1 objects that are mis-classified as Class 4 objects in the human-generated catalog. Nevertheless, the overall agreement is on-par with the human-level accuracy achieved in LEGUS and further verifies the robustness of the \starcnet model on an independent dataset from the data used for training the ML algorithm.

\begin{figure*}
  \centering
  \includegraphics[scale=0.4]{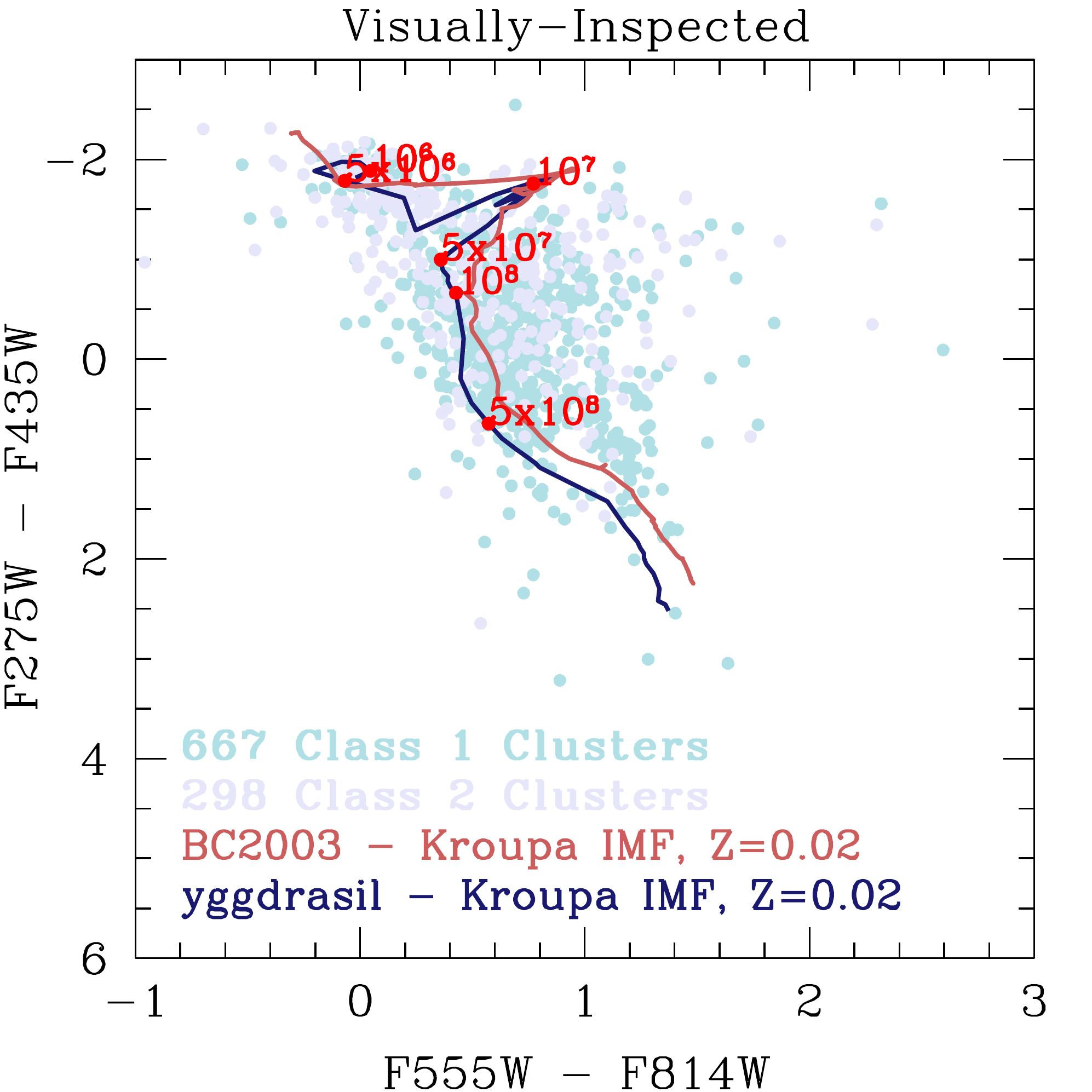}
  \includegraphics[scale=0.4]{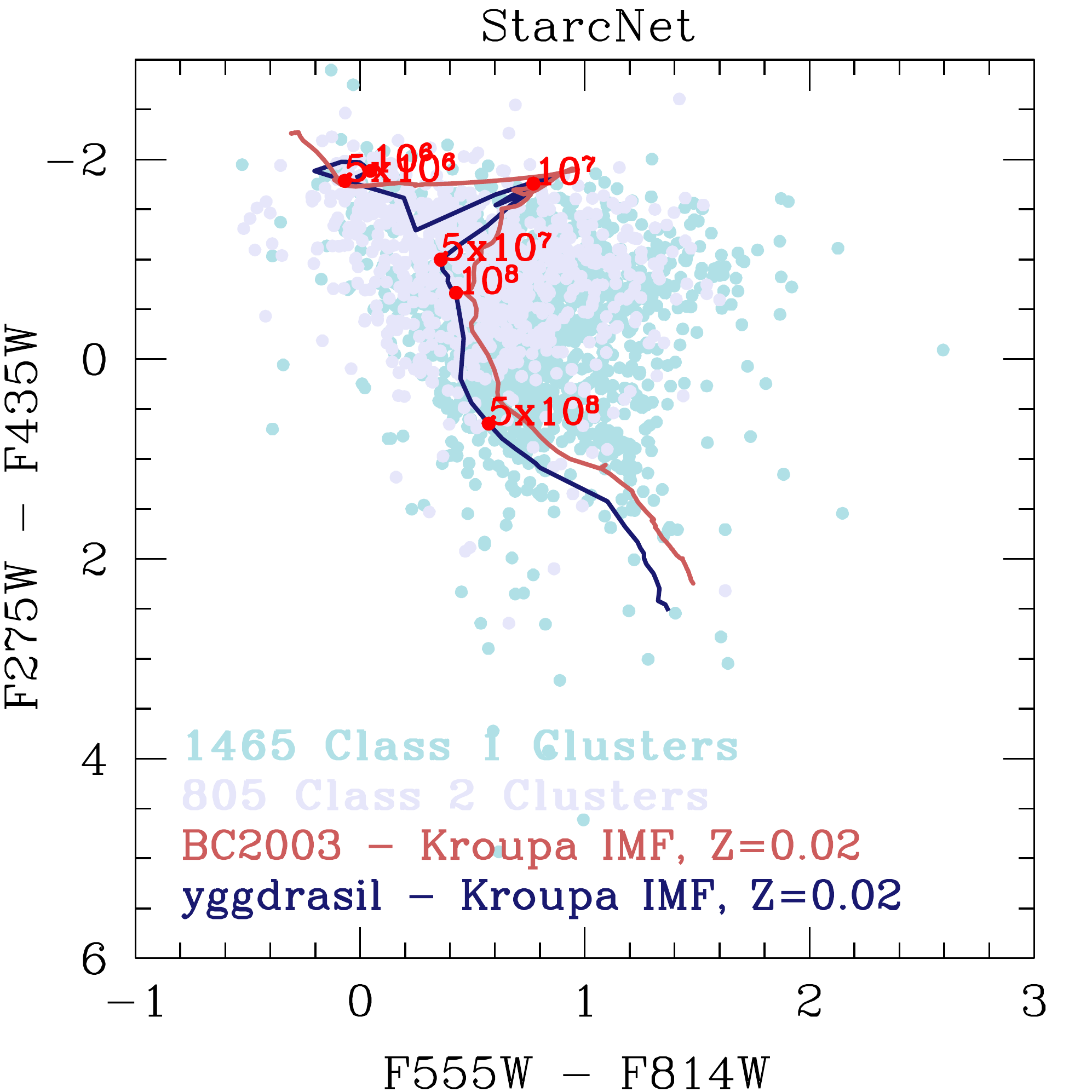}
  \caption{The NUV-B vs. V-I color diagram for all visually classified Class 1 and 2 clusters (Left Panel) and \starcnet-classified Class 1 and 2 clusters (Right Panel). Overlaid in blue and red are the yggdrasil models adopted here (with ages marked in red along the sequence), as well the comparable model from Bruzual \& Charlot (2003) without emission lines included respectively. Class 1 clusters are shown in teal with Class 2 clusters in light purple. Overall the two classes span the same range in color for both our visual and ML catalogs. The final \starcnet catalog contains $\sim 4$ times more objects than the human-generated catalog.}
\end{figure*}

In Figure 3 we show the resulting NUV-B vs V-I color-color plots for all Class 1 and 2 clusters detected for all three regions using both the human (left) and ML (right) classifications. These plots demonstrate that both catalogs span the full color range of the yggdrasil and \citet{bc03} SSP models, and thus the full range of cluster ages and extinctions, and that including ML-classified objects does not bias the distribution of cluster colors relative to the human-classified sample. The median cluster properties of each catalog along with the number of visually- and ML-identified clusters is given in Table 1, and the color-color plots for the individual regions is given in Appendix.

In particular, the median reduced $\chi^{2}$ for all \starcnet-classified Class 1 and 2 clusters detected in M101 is 1.36, with a median E (B-V) of 0.24. Importantly, we do not find any trend of increasing $\chi^{2}$ with increasing E (B-V) such that clusters with redder colors do not have systematically larger uncertainties (poorer fits) in their derived ages. Further, we do not find any significant trend between the NUV-B color and the derived cluster mass, where stochastically sampled clusters with masses of $\leq 10^{4} M_{\odot}$ may appear redder and thus older on average \citep{popescu10}. In \citet{nb11} the authors demonstrated that differences in the observed color distributions of clusters in the central- and outer-regions of M83 reflect the fact that they undergo different levels of disruption. Looking at Table 1 we see that the clusters identified in the C and SE are younger on average than the clusters in the NW which also spans a much larger breath in NUV-B vs. V-I the color distribution (Figure 13). The full analysis of the cluster age functions (CAF) in M101 are presented in \S 4.

In Figure 4 we plot the resulting distribution of Class 1 and 2 cluster V-band magnitudes in each region for both the human and \starcnet catalogs. The threshold for the magnitude cut ($M_{V} = -5.99$) was determined by using galaxies in LEGUS with low crowding and background level, and ultimately remove more objects in the C and NW region than in the SE catalog. The visually-classified clusters (red histogram) with magnitudes fainter than the cut in each region reflect objects added by-eye which were discovered in the vicinity of sources already marked for visual inspection during the classification process. However, these objects represent $\leq$ 10\% of the total sample of Class 1 and 2 clusters in the C and SE regions of M101, and span the full range of derived cluster ages. Further, we see that both catalogs return a nearly identical number of sources with apparent magnitudes brighter than $m_{V} \sim 22.5$, which suggests that the majority of real Class 1 and 2 clusters mis-classified as 3 and 4 by our visual inspection (Figure 2) are faint objects near the magnitude threshold for human-classification ($m_{V} = 23.14$) of the visually-classified sample.

\begin{figure*}
  \centering
  \includegraphics[scale=0.28]{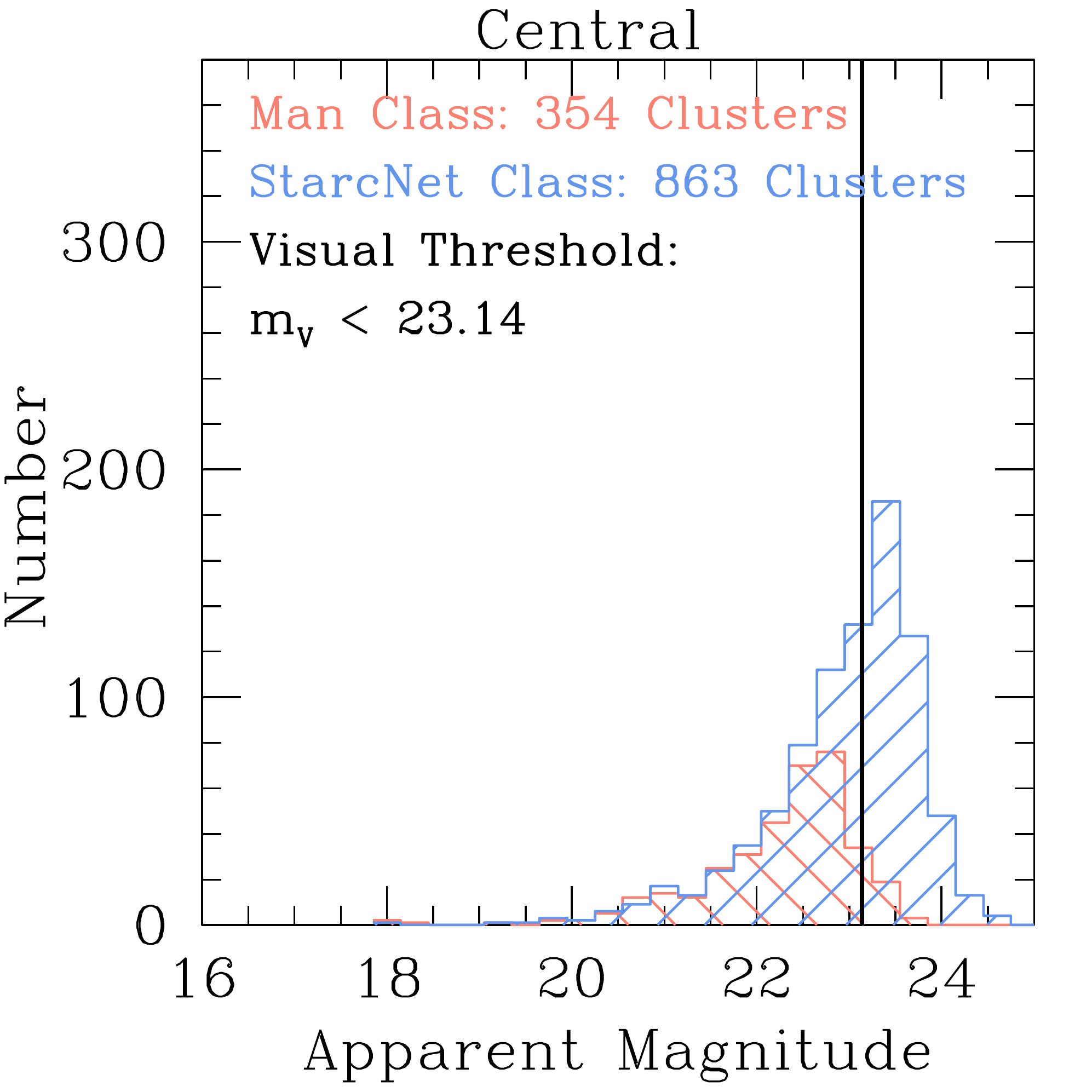}
  \includegraphics[scale=0.28]{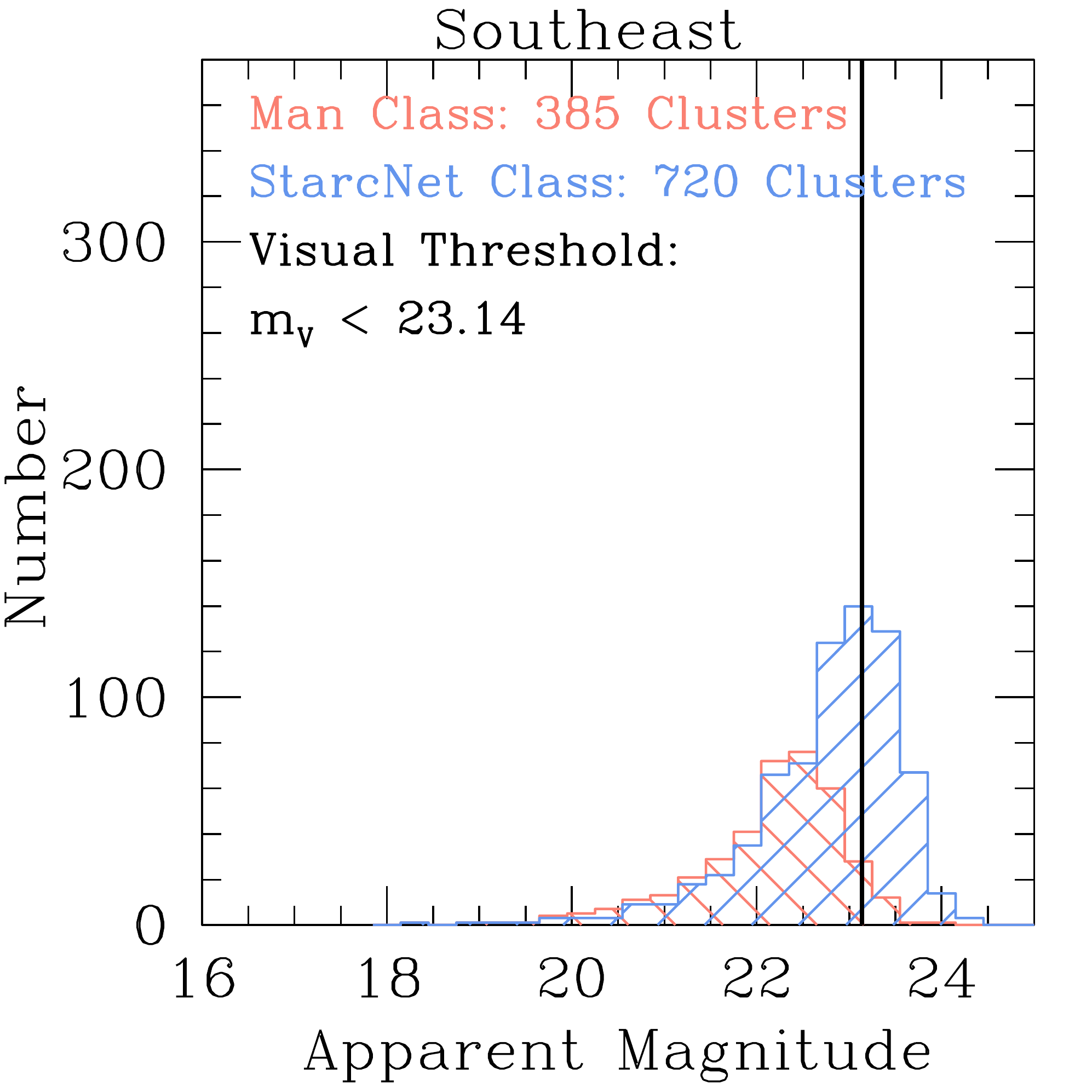}
  \includegraphics[scale=0.28]{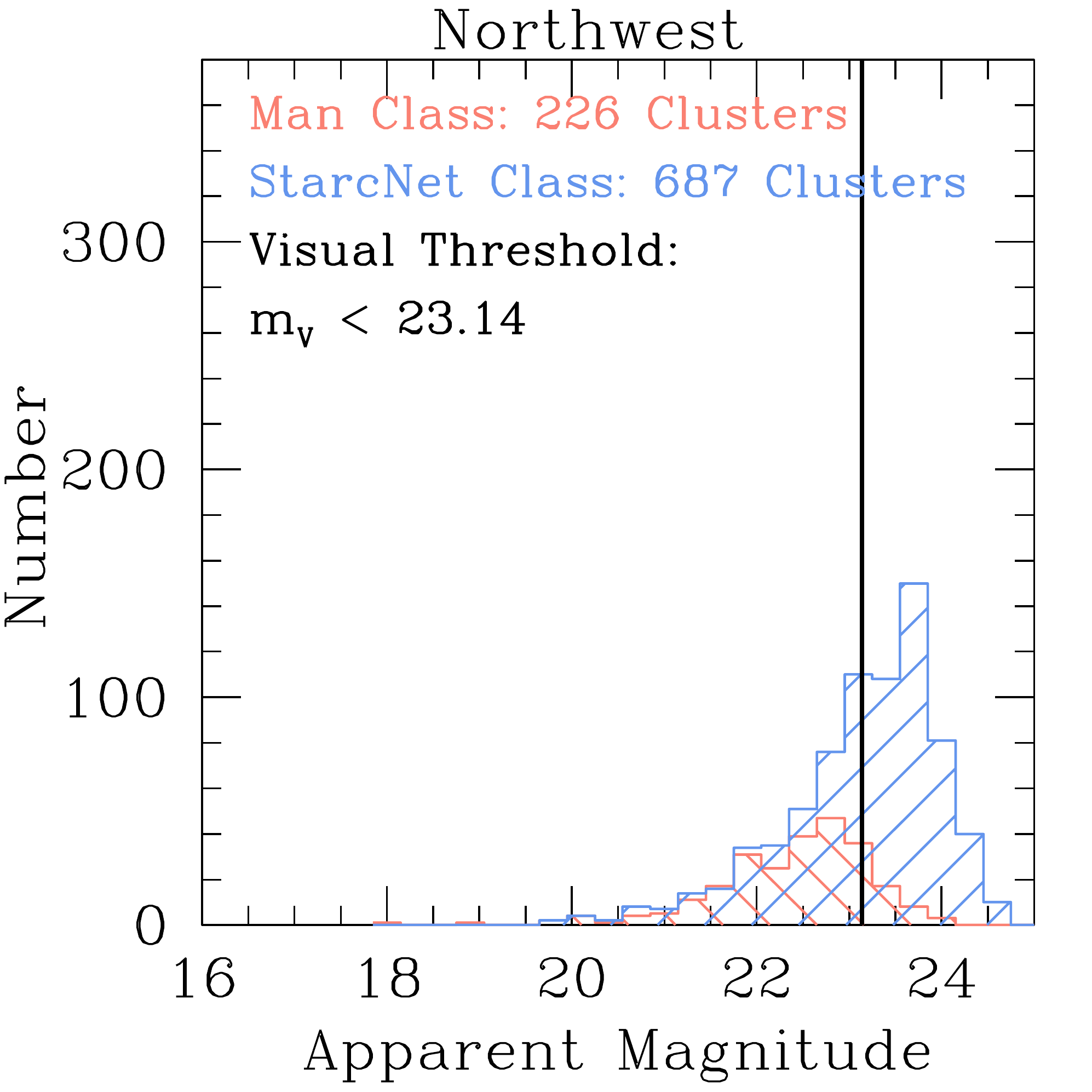}
  \caption{The distribution of V-band magnitudes in the central (left), southeast (middle), and northwest (right) regions is shown for both the visually-classified (orange) and \starcnet-classified (blue) catalogs. The magnitude threshold discussed in \S 2 for a cluster to be visually inspected ($m_{V} < 23.14$) is shown as a solid black line. The majority of the additional clusters added using the \starcnet classifications are faint, and thus crucial for extending the completeness of the sample to lower cluster luminosities.}
\end{figure*}

Finally, for all sources \starcnet returns the normalized probability distribution for the object to belong in each classification separately, where the maximum probability is adopted as the final classification. Using these probabilities we develop a negative classification index ($C_{neg} = P_{C_{3}} + P_{C_{4}}$) which is the sum probability of an object to be placed in Class 3 or 4 and quantifies the level of uncertainty each object has in being designated as a cluster vs a non-cluster in our final sample. In \S 4 we demonstrate the effects of adopting different $C_{neg}$ thresholds when analyzing the cluster age distribution for each region. Overall the final \starcnet catalog contains 2,270 Class 1 and 2 clusters, which is $\sim 4$ times larger than our final human-classified catalog of Class 1 and 2 clusters ($N_{c} = 965$), and is the catalog used to analyze the age and mass distributions of clusters in the C, SE, and NW regions of M101.

\input{tbl-1}

\section{Completeness and the Mass-Age Diagram}

\subsection{Completeness}

In order to determine the completeness limit of our cluster sample, we follow a similar prescription to \citet{aa17,mm18a} which utilizes the LEGUS Cluster Completeness Tool. We first create a set of synthetic sources with the BAOlab software \citep{larsen99} by convolving the ACS F555W PSFs with a King profile specified with effective radii of 1, 5, and 10 pc. Observations of cluster size as a function of age and mass reveal that the majority of star clusters in nearby star-forming galaxies have effective radii between 1 - 5 pc, with no clear trend between the size and the stellar mass \citep{ryon15,ryon17}. The choice of multiple cluster radii is also made to account for the fact that when the light profiles of young lower-mass clusters are generated stochastically, they may be dominated by several bright individual stars. The resulting sizes are then systematically lower than the input values which assume smoother light distributions that are well-modeled with a King profile \citep[Figure 7 - ][]{esv11}.

These synthetic sources are then distributed within our M101 scientific frames by randomly placing 1000 clusters (where $80\%$ are assigned radii between 1 - 5 pc to reflect the real populations observed in nearby galaxies) with apparent magnitudes of $18 < m_{V} < 26$ within the central, southeast, northwest 1, 2, and 3 pointings (200 clusters assigned to each). In this way we can directly account for significant variations in the depth and recovery across the disk of M101. The same extraction procedure described in \S 2 is then applied to these images and the fraction of simulated clusters recovered in the final catalog is calculated for each effective radius as a function of the input cluster magnitude. We find that after our CI cut (1.4) is applied the total V-band $90\%$ completeness limit is $m_{V} = 23.49, 23.47, 24.04$ for the C, SE, and NW catalogs respectively. \citet{aa17} demonstrated that a CI cut of 1.4 in NGC 628 only affected the recovery of very compact and partially unresolved clusters with $R_{eff} < 1$pc. Importantly, these combined limits are $\sim 0.3 - 0.5$ mag brighter than the $90\%$ completeness limits for our 1 pc synthetic sources. This ensures that our results are not biased at older ages, where cluster size increases and the luminosity decreases. Finally, we achieve a slightly better depth in the NW region where the catalog is less affected by crowding. These limits are applied to our catalogs for the following analysis of the cluster age and mass distributions.
 
\begin{figure*}
  \centering
  \includegraphics[scale=0.4]{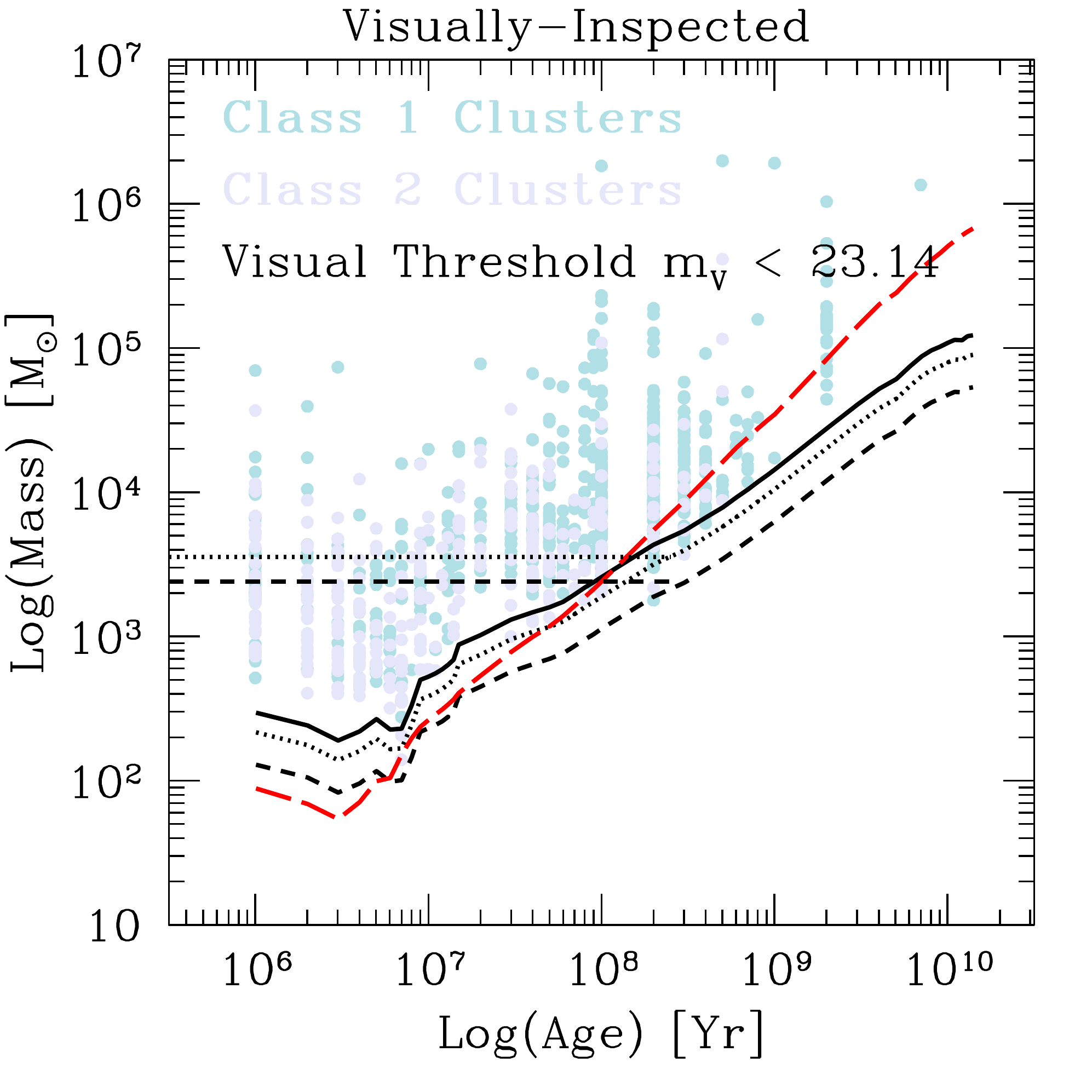}
  \includegraphics[scale=0.4]{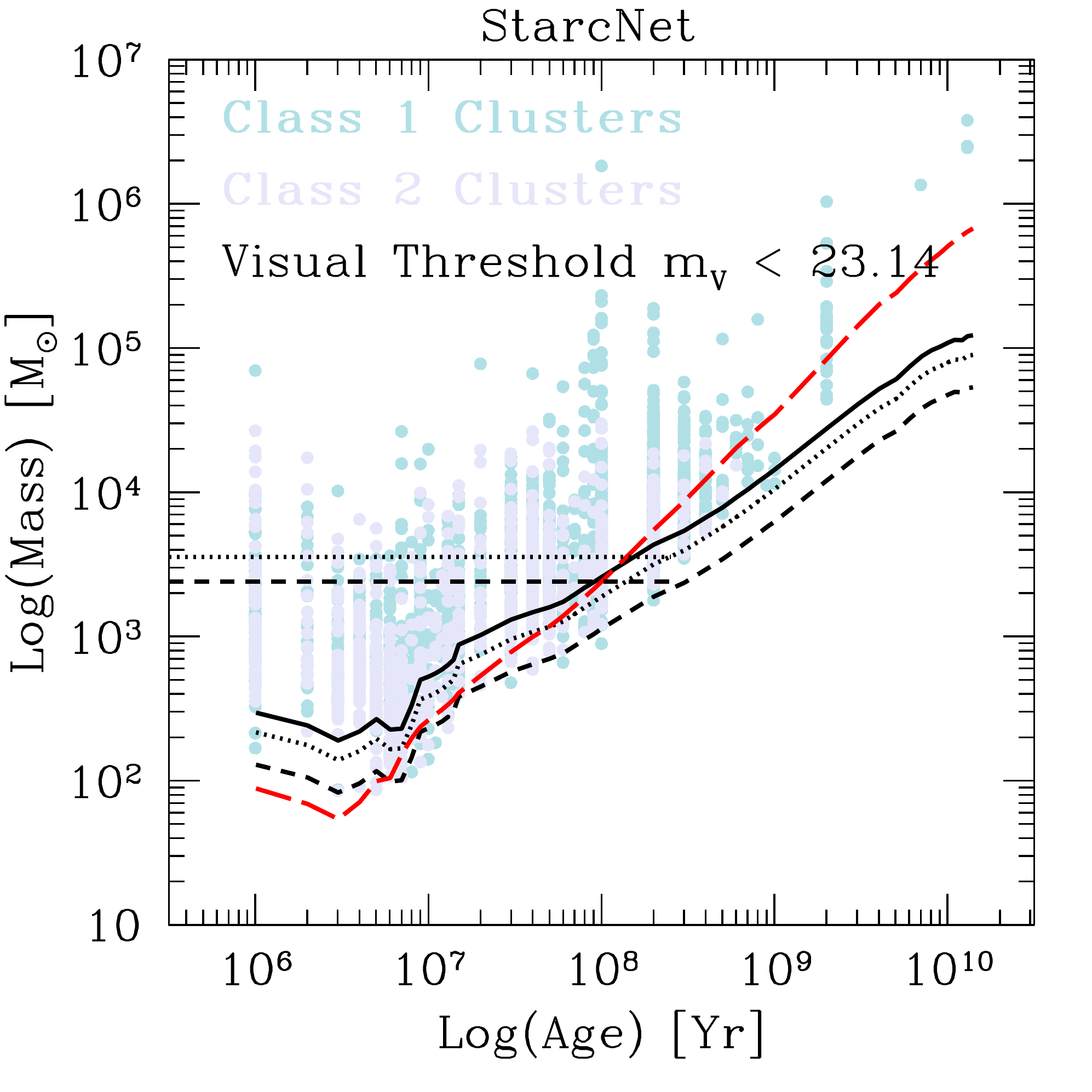}
  \caption{The age and mass for all Class 1 (teal) and 2 (light purple) clusters for both the visually-inspected (left) and \starcnet (right) catalogs. The dotted- and solid-dashed lines represent an $m_{V} = 23.48, 24.04$ respectively. The dashed-red curve represents the mass-age track for the brightest sources ($m_{U} = 23$) which begin to have U-band photometric uncertainties of $\sigma_{U} \geq 0.1$mag. All sources with $\sigma_{U} \geq 0.1$mag are removed from our final analysis. The goal of this study is to evaluate the cluster populations in M101 over the last 500 Myr. Therefore to achieve a mass-limited sample over this entire age range \citep[see][]{renaud20} we impose a final cut of $M > 10^{3.55 (3.38)} M_{\odot}$ for the C/SE (NW) regions respectively (horizontal dot and dashed lines respectively). The magnitude threshold discussed in \S 2 for a cluster to be visually inspected ($m_{V} < 23.14$) is shown as a solid black line. In addition to increasing the total number of clusters used to evaluate the age and mass distributions, the \starcnet catalog is necessary to fully capture the population of faint objects at or below the completeness limits determined for each region.}
\end{figure*}

\subsection{The Mass-Age Diagram}

Using the yggdrasil SSP we convert our $90\%$ completeness limits into tracks of age and mass. In Figure 5 we plot all Class 1 and 2 clusters for both the visually-inspected (left) and \starcnet (right) catalogs. The dotted- and solid-dashed lines represent an $m_{V} = 23.48, 23.24$ respectively. The goal of this study is to evaluate the formation and evolution of clusters in M101 over the last 500 Myr. Therefore to achieve a mass-limited sample over this entire age range \citep[see][]{renaud20} we impose a final cut of $M > 10^{3.55 (3.38)} M_{\odot}$ for the C/SE (NW) regions, shown as straight dot- and dashed black lines respectively. The dashed-red curve in Figure 5 represents the mass-age track for the brightest sources ($m_{U} = 23$) which begin to have U-band photometric uncertainties of $\sigma_{U} \geq 0.1$mag. All sources with $\sigma_{U} \geq 0.1$mag are therefore removed from our final analysis. This additional cut preferentially removes older, lower-mass sources which are not initially removed with our V-band $90\%$ completeness limits.

\citet{km15} demonstrated that Yggdrasil SSP models agree fairly well with models which stochastically sample the IMF (SLUG) at high cluster mass ($\sim \geq 10^{4.5} M_{\odot}$), but tend to be systematically low relative to the stellar masses derived with SLUG models at lower cluster mass. Despite this, the mean age and mass for the population ranging from $\sim 20$ Myr and 1.5 Gyr with $M > 10^{3} M_{\odot}$ are quite similar \citep[e.g, yellow circles in Figure 14 - ][]{km15}. Therefore we expect that at our adopted mass limit we can be reasonably confident that stochasticity has minimal impact on the sample-averaged ages and masses considered in \S 4.1.

In addition to increasing the total number of clusters used to evaluate the age and mass distributions, the \starcnet catalog is necessary to fully capture the population of faint objects at or below the completeness limits determined for each region. We also note the large number of clusters seen with ages below 10 Myr over the full range of masses. Although cluster fitting methods can create some observed structure in the mass-age diagram, e.g. a lack of clusters with ages of $\sim 15$ Myr is a common feature of model-derived ages \citep{gieles05,goddard10}, we find no evidence of such structures in either our human- or ML-generated catalogs. The mass-age diagrams for our three regions are shown individually in Appendix Figure 14.

\section{Results and Discussion: The Cluster Population in M101}

After determining ages, masses, and extinctions for all clusters across our three regions we directly compare these distributions with those of nearby normal galaxies. However before the final analysis we make three additional cuts to the cluster samples to (1) remove objects with age uncertainties of $\log (\sigma_{\tau}) > 1 $ dex despite having 4-5 filters used in the SED analysis,  and (2) for the \starcnet catalogs an additional cut of $C_{neg} < 50\%$ to ensure that no objects with a very high probability in Class 3 or 4 are retained in our final analysis. We will explore the impact of adopting different values of $C_{neg}$ in the following Section. These two cuts remove 11 and 49 clusters respectively from the human and \starcnet final catalogs.

In this Section we focus on the interpretation of the derived cluster age distributions for the \starcnet catalogs in each field relative to the underlying star formation history (SFH) in M101. We also analyze the young ($\tau < 10^{7}$ yr) and intermediate ($10^{7} < \tau < 10^{8}$) mass distribution functions in each region, and ultimately we discuss to what degree the differences observed in our cluster populations can be attributed to changes in the properties of ISM in M101. Finally, we examine the differences between the human- and ML-generated catalogs and their impact on the resulting age and mass distributions.

\subsection{Age Distributions Across the Galaxy}

In the first $\sim 100$ Myr after a gas-free cluster population emerges it is subject to mass loss via stellar evolution-driven expansion and tidal shocking by gas in its immediate
environment. The rate at which the CAF increases or declines from 10-100 Myr is parameterized by the power-law index $\gamma$, where $dN/d\tau \propto \tau^{\gamma}$. However, the slope of the CAF reflects the net difference between any increase in the star formation rate within that region over the same time interval as well as any destruction of star clusters through either a mass-independent or mass-dependent process. By jointly analyzing the SFH inferred from resolved stellar populations over the same regions in M101 we can disentangle the contributions of increased cluster formation and/or destruction to the overall CAF. 

\citet{grammer14} used archival HST ACS F435W, F555W, and F814W imaging to derive color magnitude diagrams (CMDs) for 5 annular bins in M101 starting from the center. These CMDs are then compared to synthetic models generated using the StarFISH package \citep{starfish}. These SFHs, derived using the field stars in M101, provide an independent check on the underlying star formation activity. For this analysis, we adopt Regions $\rho=1$ and $\rho=7$ in Figure 4 of \citet{grammer14} as the SFH for the C (inner) and NW (outer) regions respectively. Our SE region is blended with their $\rho=3$ and 5 annuli, which include parts of the NW disk, making it difficult to produce a SFH for the SE disk alone. The SFHs for the C and NW are then re-binned to our coarser age resolution, and fit over the same age range as our cluster age distribution with resulting slopes of $\gamma = -0.07 \pm 0.13$ and $\gamma= 0.26 \pm 0.12$ respectively.

We consider the age distribution of clusters in our three M101 regions from $10^{7}$yr $< \tau < 10^{8.5}$yr. This choice allows us to make accurate comparisons to the age distributions of clusters in other nearby galaxies, while also avoiding selecting the youngest objects, many of which may be unbound \citep{kab16, rc15, aa20a, bag21}. Further, when analyzing the cluster age distributions we exclude the largest mass bins of $\log (M/M_{\odot}) > 7.0$. These very high masses may be the result of either an imperfect extinction correction or multiple very compact star clusters in close proximity appearing as a single star cluster at the resolution of these images. 

Several recent studies of star clusters in nearby galaxies have noted that when adopting a single metallicity value for SED fitting, many old massive globular clusters may appear artificially as younger ($1-5$x$10^{8}$yr) more heavily-extincted objects \citep{bcw20}. Because these clusters are all high-mass, their mis-classification could potentially bias the interpretation of the age distribution slopes over this age regime \citep[e.g.,][]{deger22,hannon22,moeller22}.~Although we stress that sources with large uncertainties in their derived ages ($> 1$ dex) are already removed from this analysis, we find 19 sources with ages $10^{7.8}-10^{8.4}$yr and with masses $> 10^{4.8} M_{\odot}$. To assess the impact of a 'worst case' scenario where all of these objects are indeed mis-classified globular clusters, we remove them from our analysis of the age distributions and find that the changes in the measured slopes are consistent within the uncertainties of the fit.

\begin{figure*}
  \centering
  \includegraphics[scale=0.28]{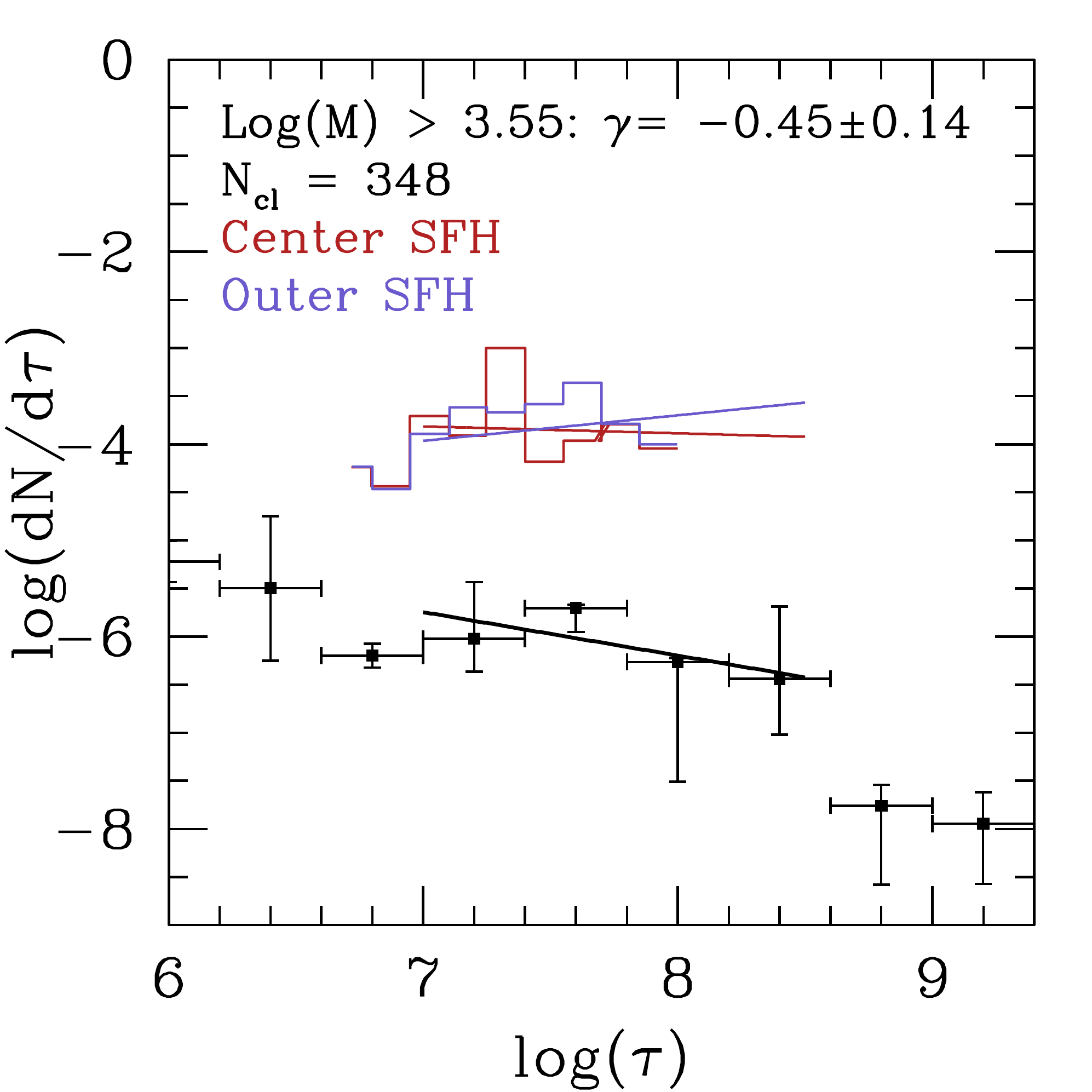}
  \includegraphics[scale=0.28]{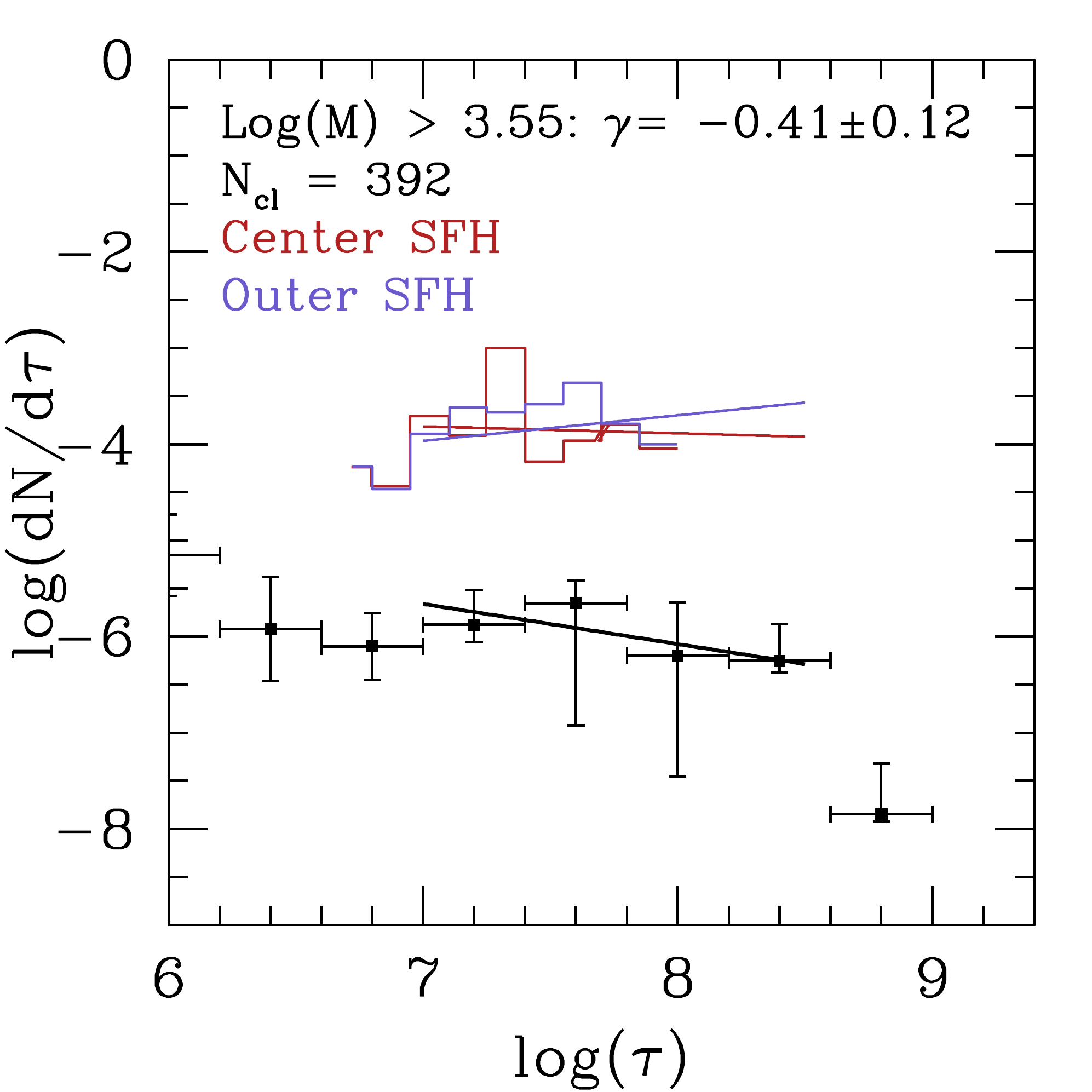}
  \includegraphics[scale=0.28]{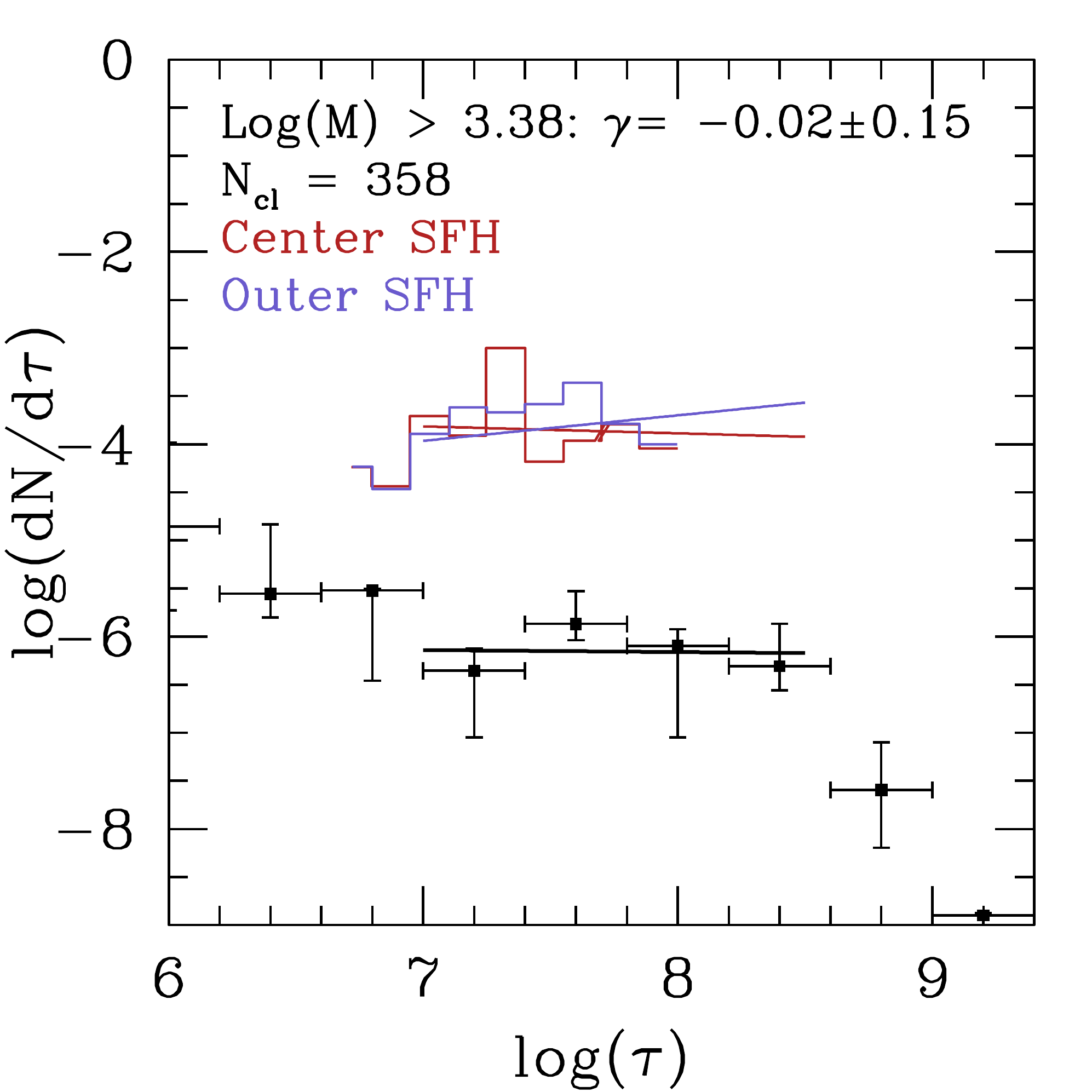}
  \caption{The differential number of clusters per age bin, $\log (dN/d\tau)$, versus the median cluster age in each bin for all clusters with $10^{3.55 (3.38)} < M/M_{\odot} < 10^{7}$ in the C (left), SE (middle) and NW (right) regions respectively. The total number of Class 1 and 2 clusters (determined by \starcnet) which are ultimately used for this analysis is given in the upper-left of each Panel. We overlay the CMD-based SFHs for the C (red) and NW (purple), which are re-binned and fit over the same age range as our cluster age distributions, demonstrating that the negative slopes measured for $\gamma$ are indeed due to cluster disruption over the mass and age ranges considered \citep{grammer14}.}
\end{figure*}

In Figure 6 we show the differential number of clusters per age bin, $\log (dN/d\tau)$, versus the median cluster age in each bin for all clusters with $10^{3.55 (3.38)} < M/M_{\odot} < 10^{7}$ in the C (left), SE (middle) and NW (right) regions respectively. The plotted data are binned by 0.4 dex in $\log (\tau)$ so as to fully encapsulate $\sim 3\sigma$ times the typical model errors of 0.2 in $\log (\tau)$ discussed in \S 3. We overlay the inner (red) and outer (purple) SFHs for each region, shifted to lie on top of one another for comparison in Figure 6. It is clear that our measurements of $\gamma \leq 0$ for all three regions are not driven by a similarly negative slope in the CMD-based SFHs. In fact, the CMD data suggests the SFH in the C and NW regions of M101 has been roughly constant or even decreasing in the center over the last 300 Myr. These results are also consistent with a recent pixel-based panchromatic SED analysis of the SFH for 10 nearby galaxies, finding that the specific star formation rate and star formation efficiency are suppressed in the C region of M101 despite the increase in the fraction of molecular gas \citep{pixsed22}. Thus, we can reasonably interpret the steepness of the age distributions as reflective of the strength of cluster disruption mechanisms in that region.

\begin{figure}
\centering
  \includegraphics[scale=0.4]{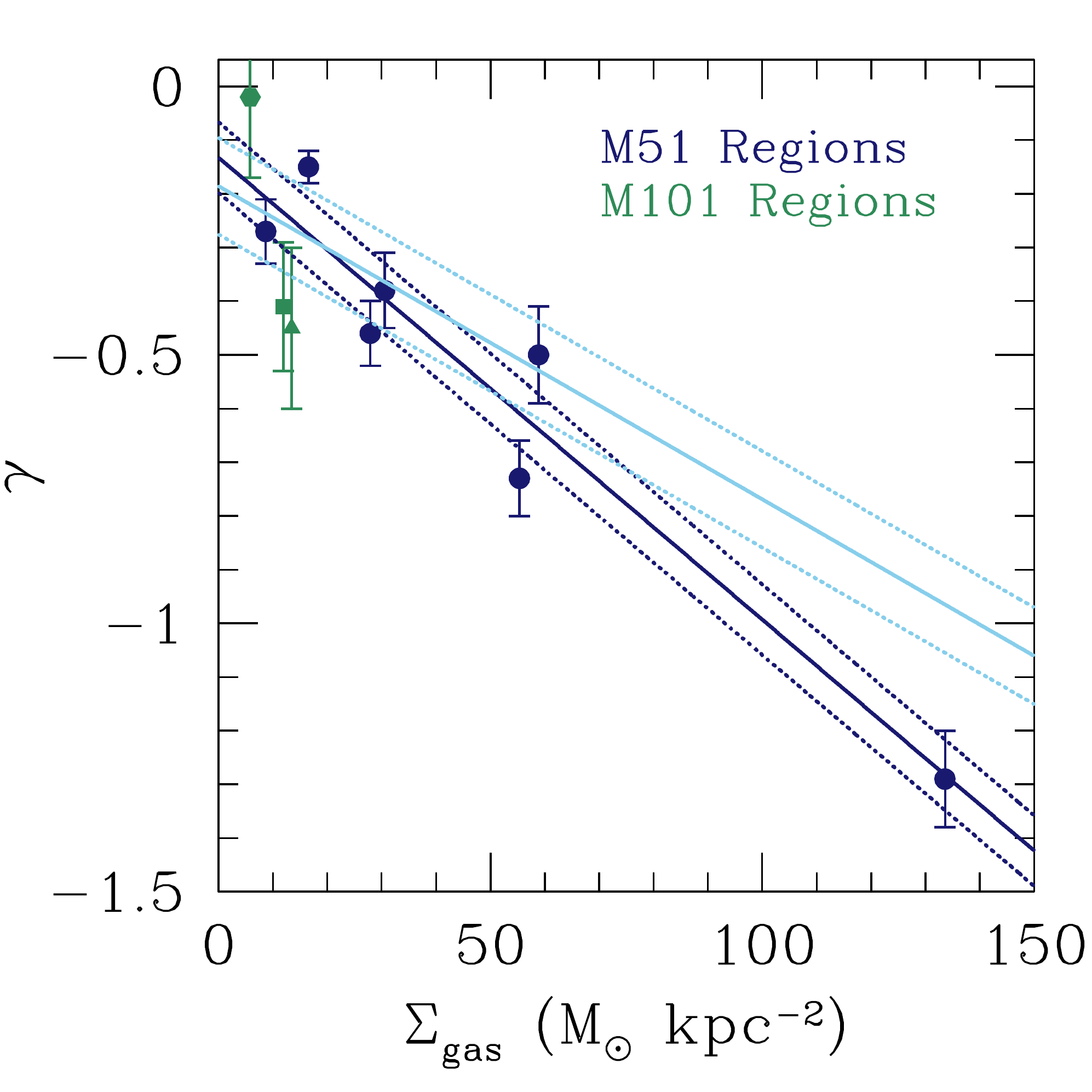}
  \caption{Values for the cluster disruption rate ($\gamma$) and the molecular gas surface density ($\Sigma_{gas}$) over the same field-of-view plotted for the C (filled triangle), SE (filled square), and NW (filled Hexagon) regions in M101 (green) and regions in M51 (filled blue circles) taken from \citet{mm18b}. A linear least-squares fit to the M51 data points along with the corresponding 1$\sigma$ uncertainties are shown in blue. The fit excluding the regions with the highest disruption in M51, which correspond to the molecular ring (MR) and spiral arms (SA), is shown by the light blue line. The central region of M101 appears to have a slightly higher disruption rate for the measured $\Sigma_{gas}$ relative to both of the trends inferred from the M51 data.}
\end{figure}

To this end, we find that the difference in the measured $\gamma = -0.45 \pm 0.14$ and $\gamma = -0.02 \pm 0.15$ between the C and the NW is statistically significant. Further, given that the SFH of the outer-disk of M101 has a positive CAF slope ($\gamma = 0.26$) the observed difference may actually be a lower-limit. This trend is also similar to what has been found in M83 and M51 \citep{esv14,mm18b}, although the overall cluster disruption appears to be stronger in the central region of M51. This may reflect the larger star formation and gas surface densities seen as a function of decreasing galactocentric radius in M51 \citep{querejeta19}. Using the gas-to-dust (G/D) ratio maps of M101 in \citet{sandstrom13} we derive a gas surface density for the C and NW regions of M101 to be $\Sigma_{gas} \sim 15 M_{\odot}$pc$^{-2}$ and $\Sigma_{gas} \sim 4-6 M_{\odot}$pc$^{-2}$ respectively depending on the exact choice for the CO-$H_{2}$ conversion factor $\alpha_{CO}$ used. These values are consistent with GALEX and \textit{Spitzer} measurements of the total SFR (UV+24$\mu$m) for each region in M101 converted to molecular gas with the resolved Kennicutt-Schmidt relation (rKS), for which a recent analysis from PHANGS determined to be $\Sigma_{gas} \propto \Sigma_{SFR}^{1.01 \pm 0.01}$ at 100 pc scales \citep{pessa21}.

To compare our results directly with other studies of nearby galaxies we adopt the values of $\Sigma_{gas}$ and $\gamma$ from Figure 14 of \citet{mm18b} for 7 different morphologically-defined regions in M51: the molecular ring (MR), inter-arms (IA), spiral arms (SA), and 4 radial bins each containing an equal number of clusters which extend from the center to $\sim 7$ kpc \citep[see Figure 1 of][]{mm18b}. We then perform a linear least-squares fit to the regions in M51 to determine how the gradient in molecular gas surface density compares to our results for the C, SE, and NW regions of M101 (solid blue line). In Figure 7 we see that the overall disruption in M101 is lower compared to the inner spiral arms and molecular ring in M51, and the only regions with comparably low $\Sigma_{gas}$ values to M101 are the outer-most radial bin and the inter-arms of the galaxy. However, the central and SE regions appear to have a slightly higher disruption ($\sim 2\sigma$) for their $\Sigma_{gas}$ relative to the linear least squares fit of the M51 data. 

Differences in the completeness limits between the two cluster catalogs may artificially increase the strength of the correlation observed. From \citet{mm18a} we see that the V-band completeness limit for the outer-disk is comparable to the value found for the NW regions of M101. However for the central region of M51, which has a much higher measured $\Sigma_{SFR}$ and $\Sigma_{gas}$ \citep{schuster07,schinnerer13}, the completeness limit is $10^{3.7} M_{\odot}$. Adopting this value for the C region of M101 does not significantly flatten the observed age distribution slope within the uncertainties. Finally, even after excluding the regions in M51 with the steepest age distribution slopes (MR and SA) and re-fitting the correlation (light blue line), the age distribution slope the central region of M101 remains steeper. It is therefore likely that neither the trend we observe between $\gamma$ and $\Sigma{gas}$ or the differences between the two galaxies are driven by incompleteness.

As demonstrated in \citet{linden21}, changes in the observed disruption rate of star clusters in the central regions of major galaxy mergers are broadly consistent with models where large increases in the density of gas in the ISM can cause clusters to be destroyed through collisions with GMCs over timescales of 200-300 Myr \citep{miholics17}. An additional mechanism that can affect the survival times of GMCs and star clusters is destruction due to galactic shear. Using the EMOSAIC simulations of cluster formation and evolution across cosmic time \citet{jeffreson20} demonstrated that for MW-like galaxies galactic shear can be the dominant mechanism of GMC destruction, with a characteristic lifetime of $\sim 200$ Myr, for values of mid-plane pressure $P_{mp} k_{B}^{-1} > 10^{4.2}$ K cm$^{-3}$ which is only achieved in the central region of M101. Therefore a factor of 3 increase in the gas surface density from the outer to the inner regions of M101, as well as an additional contribution from an increased galactic shear may account for the differences in the measured slope $\Delta \gamma \sim 0.4$ dex.

\subsection{Age Distributions Over Different Mass Ranges}

\begin{figure}
\centering
  \includegraphics[scale=0.35]{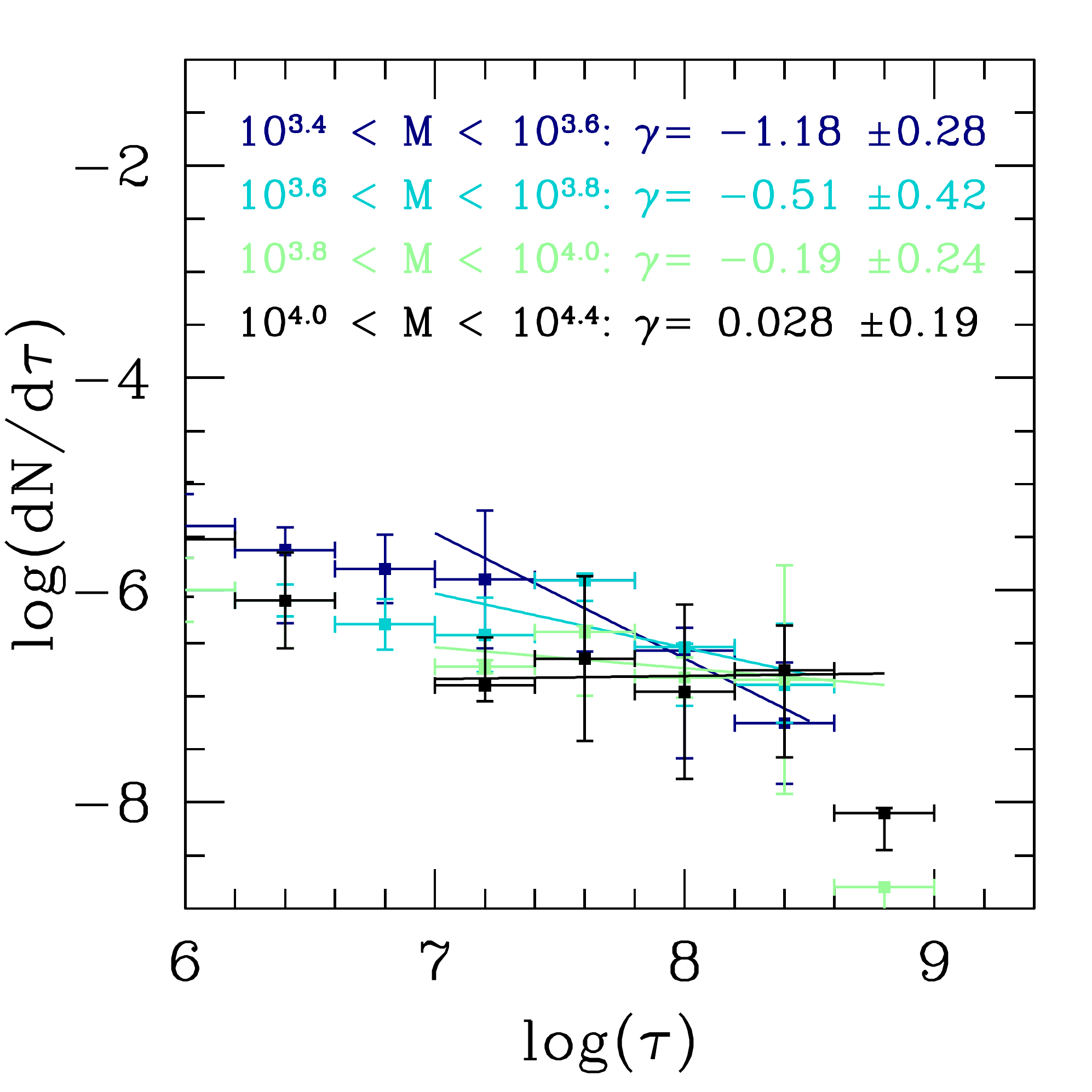}
  \includegraphics[scale=0.35]{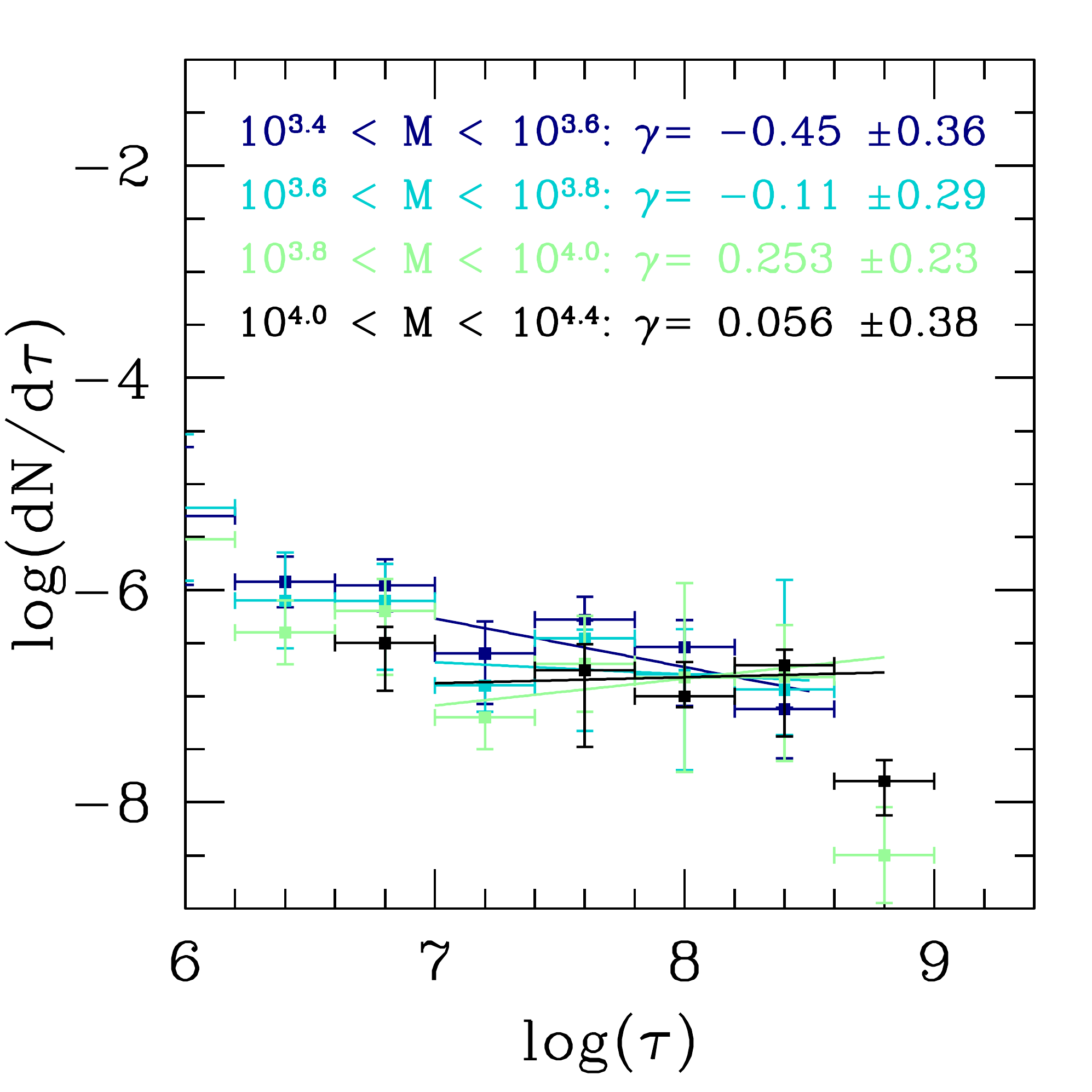}
  \caption{The age distributions for Class 1 and 2 clusters in the C (top) and NW (bottom) fields. In order to test for a mass-dependence on the measured slope $\gamma$ we increase the mass rage of the clusters considered from $10^{3.55} < M/M_{\odot} < 10^{3.6}$ (dark blue), $10^{3.6} < M/M_{\odot} < 10^{3.8}$ (light blue), $10^{3.8} < M/M_{\odot} < 10^{4.0}$ (light green), and $10^{4.0} < M/M_{\odot} < 10^{4.4}$ (black) showing that lower-mass clusters experience a higher rate of overall disruption relative to low mass clusters within the same region of M101.}
\end{figure}

Finally, in order to determine whether the results for our age distribution slope is dependent on the range of cluster masses considered we re-measure $\gamma$ for the C, SE, and NW regions using mass intervals of $10^{3.4} < M/M_{\odot} < 10^{3.6}$, $10^{3.6} < M/M_{\odot} < 10^{3.8}$, $10^{3.8} < M/M_{\odot} < 10^{4.0}$, and $10^{4.0} < M/M_{\odot} < 10^{4.4}$. In Figure 8 we show the resulting age distribution slopes for the C and NW regions. For both regions we find that increasing the mass range considered has the effect of flattening the observed distribution slope, which is tentative evidence for mass-dependent cluster disruption. However, given the uncertainties in estimating the completeness limit at the faint-end, we require comparably significant differences in the slopes of the CMFs between different regions to claim evidence for this effect (see \S 4.3). Importantly, the distribution slopes measured for the C region also remain steeper than the NW regardless of the mass range considered. These conclusions are what we would expect for environmentally-dependent disruption for the cluster populations in M101, and are similar to the results found when analyzing the young and old cluster populations in M82 \citep{li15}. In Table 2 we give the number of clusters used to fit the distributions shown in Figure 8 for all three regions in M101.

\input{tbl-2}

\subsection{Mass Function}

The cluster mass function (CMF) has been studied extensively across a wide variety of star-forming galaxies in the local Universe. These studies have produced a canonical CMF slope of the form $dN/dM \propto (M)^{-2}$ \citep[see][]{aa20b}. We consider the mass distribution of clusters in M101 over two age intervals to sample the young ($\tau < 10^{7}$) and intermediate ($10^{7} < \tau < 10^{8}$) age populations. When modeling the CMF we adopt a maximum-likelihood fitting technique as well as a bootstrap resampling of the posterior distribution functions for a simple model (PL) model where $\beta$ is the overall power-law slope. We implement the IDL routine MSPECFIT, which has been used previously to study the mass functions of both giant molecular clouds and YSCs in galaxies in the local Universe \citep{rosolowsky07,mm18a}. A study to look for evidence of a mass truncation in the CMFs measured for the full LEGUS survey is the topic of an upcoming analysis (Adamo et al. in prep), and is outside of the goals of this paper.

\begin{figure*}
  \centering
  \includegraphics[scale=0.4]{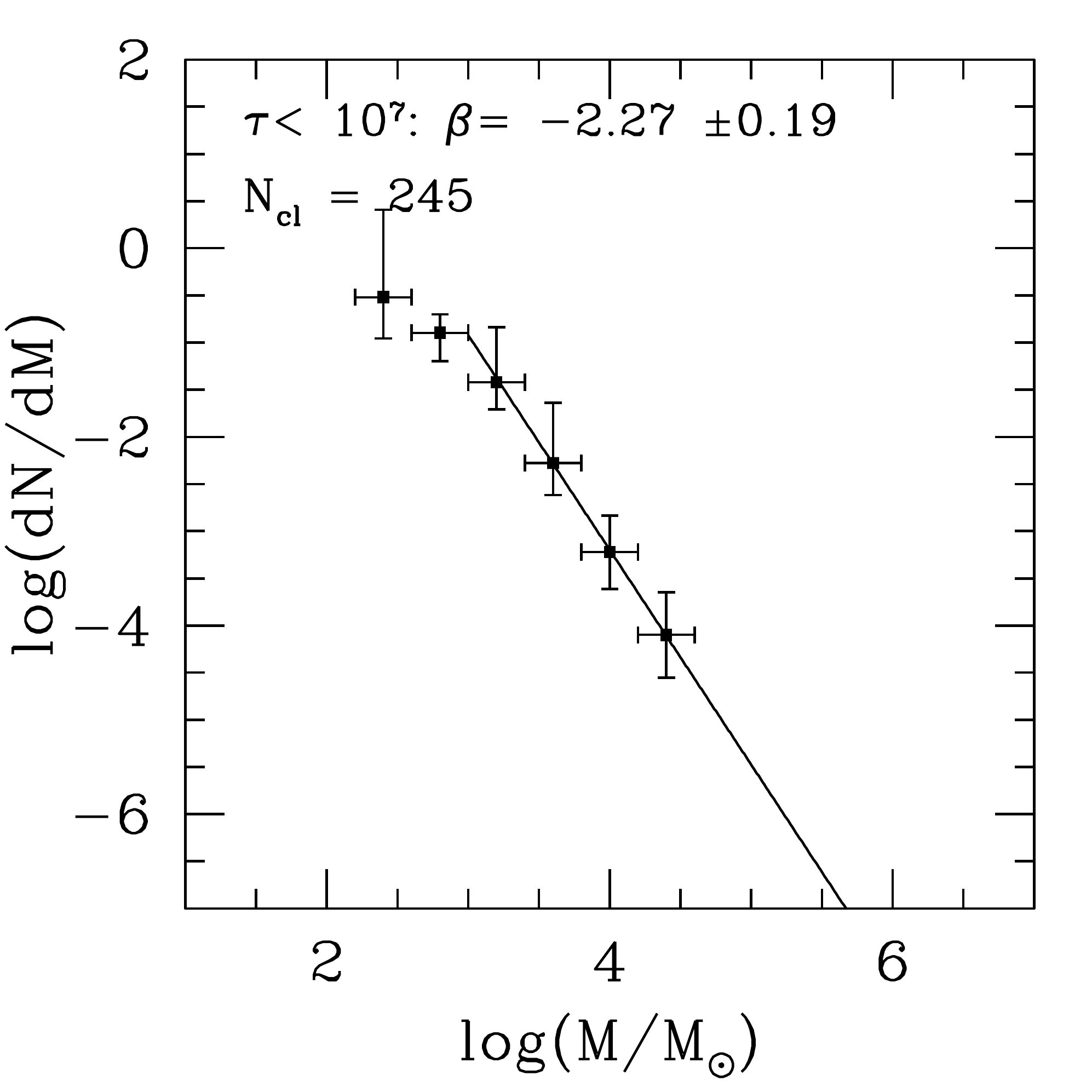}
  \includegraphics[scale=0.4]{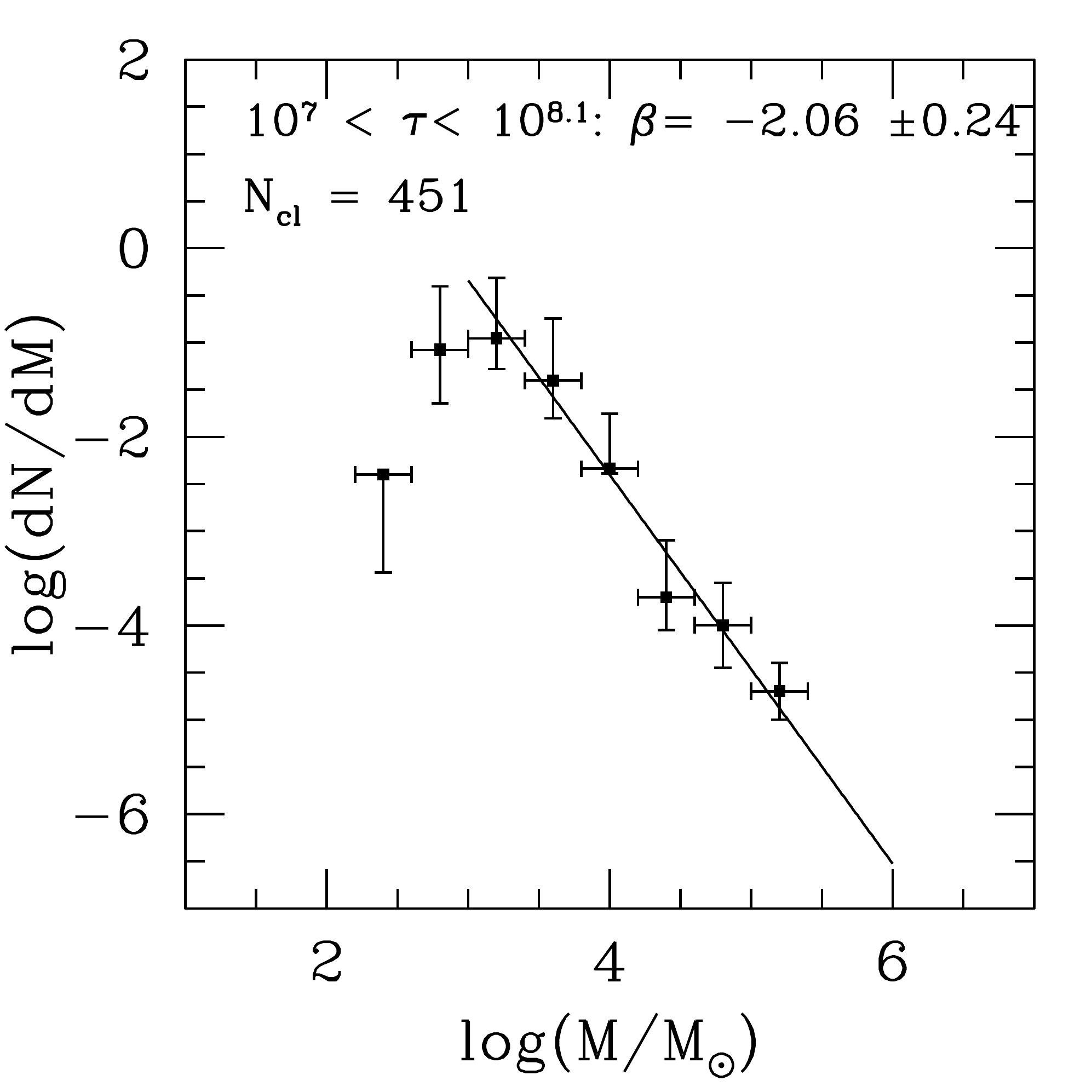}
  \caption{The cluster mass function (CMF) for young ($\tau < 10^{7}$ yr - left Panel) and intermediate-age ($10^{7} < \tau < 10^{8}$ - right Panel) star clusters in our \starcnet catalog for the central pointing. Both distributions are well-fit with a power-law $dN/dM \propto M^{\beta}$, where $\beta$ is consistent with the canonical $\sim -2$ PL slope for both age intervals.}
\end{figure*}

In Figure 9 we show the young and intermediate age CMFs for the C region in M101 using our \starcnet catalog. We find that the data are well-described by a power-law with a slope $\beta \sim -2$ over both age intervals within uncertainties. Further, the CMFs derived using the human-classified catalog for the C region are $\beta_{young} = 2.27 \pm 0.19$ and $\beta_{int} = -2.06 \pm 0.24$. These vales are consistent within uncertainties and again suggest that the ML algorithm does not preferentially mis-classify clusters with lower-luminosities. A similar result was also found by comparing the CMF for human-classified and machine-learning classified catalogs in M51 \citep{grasha19}. We also note that these results are obtained using the same mass completeness limit of $10^{3.55} M_{\odot}$ for both the human-classified and ML cluster samples. In Table 3 we list the derived CMF slopes for each age interval, and for each region based on the \starcnet catalogs. In Figures 17 and 18 we show the CMFs derived for the young and intermediate-age clusters in the SE and NW catalogs respectively.

In Figure 8 we found tentative evidence for mass-dependent cluster destruction in both the C and NW fields, where the overall disruption is significantly higher in the center of M101 relative to the outer-disk. If both environmentally- and mass-dependent cluster destruction are occurring in M101, we expect to see a comparably significant difference in the CMFs between the C and NW regions overall as well as between the young and intermediate age cluster samples within a single region. In Figure 9 we see find that the young and intermediate CMF slopes are consistent within uncertainties, however the low-mass turnover increases by $\sim 0.5$ dex from $10^{2.5} M_{\odot}$ to $10^{3.0} M_{\odot}$. In contrast, the low-mass turnover in the NW increases by a smaller amount, $\sim 0.25$ dex from $10^{2.5} M_{\odot}$ to $10^{2.75} M_{\odot}$, between the young and intermediate age cluster populations respectively. Further, unlike the C and SE regions the slope of the CMF in the NW appears to be slightly steeper, at the $\sim 2\sigma$ level, than a canonical $-$2 power-law for both the young- and intermediate-age cluster samples (Appendix Figures 15 and 16). While, these differences hint at possible mass-dependency in the overall disruption rate between the C and NW regions of the galaxy we stress that determining the low-mass turnover in the CMF for cluster masses $M<10^{3} M_{\odot}$ may be significantly affected by stochasticity, and thus we do not find strong evidence in support of mass-dependent cluster disruption in M101.

\input{tbl-3}

\subsection{Comparing to Visually-Selected Cluster Samples}

An open question in our analysis remains: by adopting the \starcnet catalog in the previous subsections what are the overall effects on the cluster color distributions, and the subsequently-derived cluster age and mass distributions? In Figure 10 we first show the color-color distribution of all Class 1 and 2 clusters in our \starcnet catalogs separated into 4 bins based on their apparent magnitude: $m_{V} < 23.14$, $23.14 < m_{V} < 23.48$, $23.48 < m_{V} < 24.04$, $24.04 < m_{V}$. These bins represent clusters brighter than our threshold for visual classification, clusters brighter than the 90$\%$ completeness limit for the C and SE regions, clusters brighter than the 90$\%$ completeness limit for the NW region, and clusters fainter than this limit which are not used in any analysis. We can see that the median cluster colors for the first three magnitude bins are approximately the same. It is not until we look at clusters fainter than $m_{V} = 24.04$ do we see the median value (black square) shifts to redder colors, and thus this magnitude bin does not represent the full distribution of cluster colors determined from the brightest sources in the catalog. 

\begin{figure*}
\centering
     \includegraphics[scale=0.6]{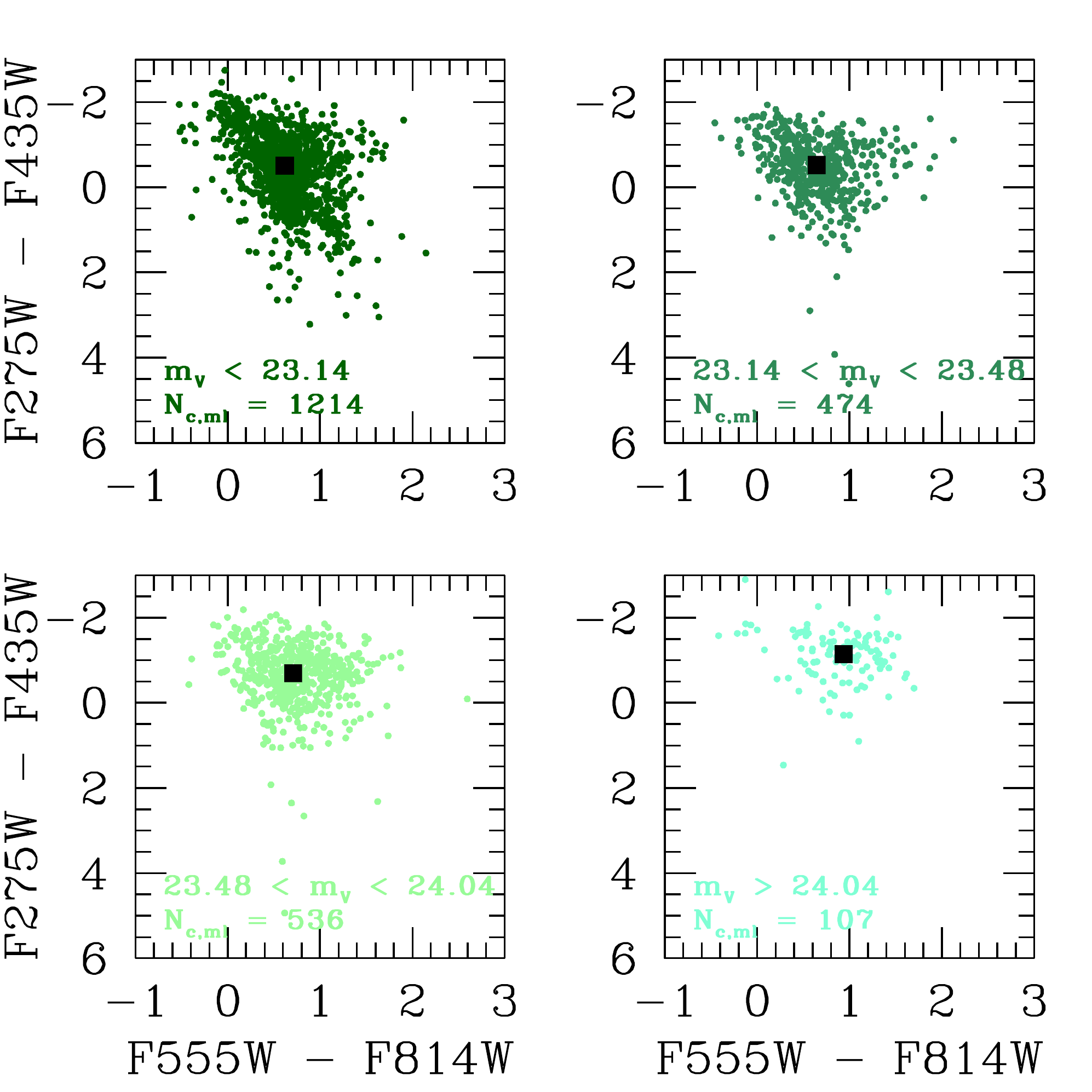}
\caption{The NUV-B vs. V-I color diagram for all visually classified Class 1 and 2 clusters in 4 bins based on their apparent magnitude: $m_{V} < 23.14$ (top left), $23.14 < m_{V} < 23.48$ (top right), $23.48 < m_{V} < 24.04$ (bottom right), $24.04 < m_{V}$ (bottom left). These bins represent clusters brighter than our threshold for visual classification, clusters brighter than the 90$\%$ completeness limit for the C and SE regions, clusters brighter than the 90$\%$ completeness limit for the NW region, and clusters fainter than this limit which are not used in any analysis. Overall the clusters used in our analysis $m_{V} < 24.04$ span the full range in cluster colors. The median color in each bin is shown by the back square in each panel.}
\end{figure*}

To further investigate the nature of the faint sources included in our \starcnet cluster catalogs we compare the positions of Class 1 and 2 clusters to the population of faint red clusters discovered in the disk of M101 with previous HST ACS imaging \citep{simanton15}. These clusters are one magnitude fainter than the expected GC peak ($M_{V} = 6.5$) in M101 based on observations of the GC luminosity function in the MW. They occupy the same luminosity-color space as LMC intermediate age clusters, suggesting that these objects have survived a long time in the disk without being disrupted. This scenario may be consistent with the lower disruption rates seen for the three regions in M101 relative to observations of other nearby spiral galaxies \citep{mm18b}. Over the WFC3+ACS footprint of our study \citet{simanton15} discovered 140 such clusters, and we retain 101/140 and 132/140 of these objects in the human-generated and \starcnet-generated catalogs respectively. This again demonstrates that the faint cluster population in our \starcnet catalog is not biased towards sources of a particular color ranges.

To assess the impact on the subsequently-derived cluster age and mass distributions we quantify how the human-generated and \starcnet classifications depend on the physical properties of the star clusters. Overall we find that the agreement remains high ($\sim 78\%$) relative to the full sample (Figure 2) for Class 1 and 2 clusters with $10^{3.38} < M_{\odot} < 10^{4.5}$ and rises above $\sim 90\%$ for Class 1 and 2 clusters with $10^{5} < M_{\odot} < 10^{7}$. Further, when comparing the classifications as a function of cluster age we again find good agreement ($\sim 81\%$ and $\sim 72\%$) for Class 1 and 2 clusters with ages 1-100 Myr and 100-1000 Myr respectively. These differences do not significantly affect our interpretation of the results (environmentally-dependent cluster disruption) due to the fact we see these trends across all three fields in M101.

Several reviews \citep{krumholz19,aa20b} have suggested that inclusive source selection (automatic detection and classification) vs. exclusive (catalogs which have been visually-inspected) may be a factor which has led different research groups, producing catalogs in different ways, to reach different conclusions. In Figure 11 we show the cluster age distribution function for the central region in M101 produced using the visually-inspected catalog as well as the same completeness limits in mass and age used in Figure 6. We find that the inferred distribution slope is consistent within uncertainties relative to the \starcnet catalog for the same region and over the same age and mass ranges.

\begin{figure}
\centering
     \includegraphics[scale=0.4]{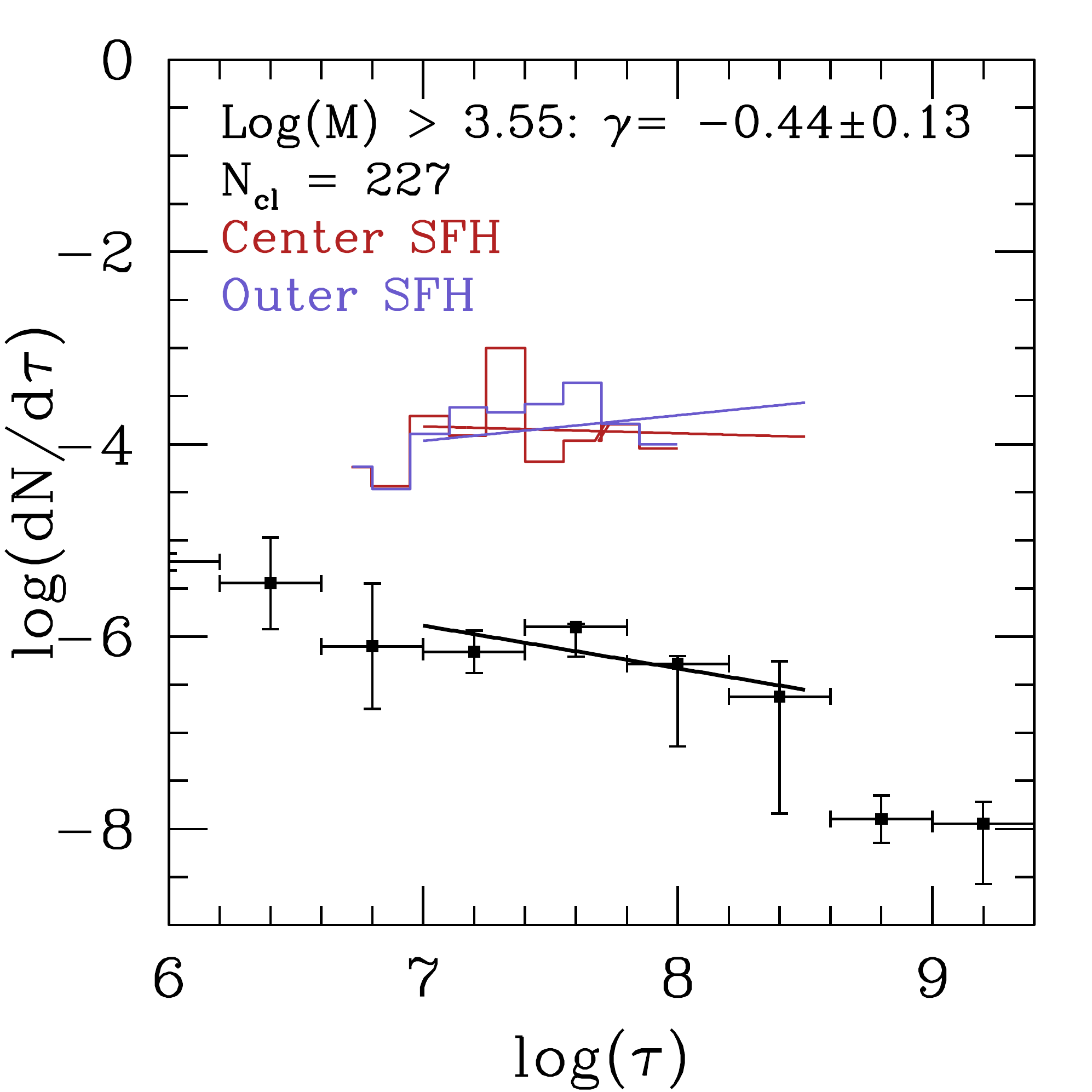}
\caption{The age distributions for Class 1 and 2 clusters in the C field using our visually-classified catalog. The distribution slope appears to be consistent with the \starcnet catalog, and underscores the fact \starcnet is capable of producing a catalog of low- and high-mass objects with robust classifications in M101 without out the need for human-inspection.}
\end{figure}

Thus, although human-classified catalogs have generally been preferred in the literature, with the inclusion of \starcnet we can produce a catalog of low- and high-mass objects with robust classifications without the need for time-intensive human-inspection; this time-intensive process fundamentally limits our ability to characterize the complete cluster populations in the most nearby massive spiral galaxies.

Finally, in order to determine the effect that the \starcnet Class 1 and 2 confidence thresholds have on the resulting distribution we show the age distribution functions for the C region of M101 with 4 different thresholds of $C_{neg}$ in Figure 12. Along with the number of Class 1 and 2 clusters included within each threshold in the top left of each Panel, we also give the resulting slope $\gamma$. As we increase the threshold for including in the cluster catalog from $C_{neg} < 50\%$ to $C_{neg} < 20\%$ the slope begins to flatten, though remains consistent within the uncertainties of the measurements. Thus, while higher thresholds for including a cluster in the \starcnet catalog reduces the overall number of objects, this does not preferentially remove only low-mass clusters from the final sample. In the Appendix Figures 17 and 18 we show the corresponding age distributions for the SE and NW regions. A similar flattening is observed for these catalogs, but importantly the difference between the disruption rate measured for the inner and outer-regions of M101 $\Delta \sim 0.2$ is robust against different thresholds of $C_{neg}$, and further demonstrates that the inclusion of \starcnet classifications does not introduce any systematic biases in the resulting cluster catalogs or derived age distributions.

\begin{figure*}
\centering
     \includegraphics[scale=0.6]{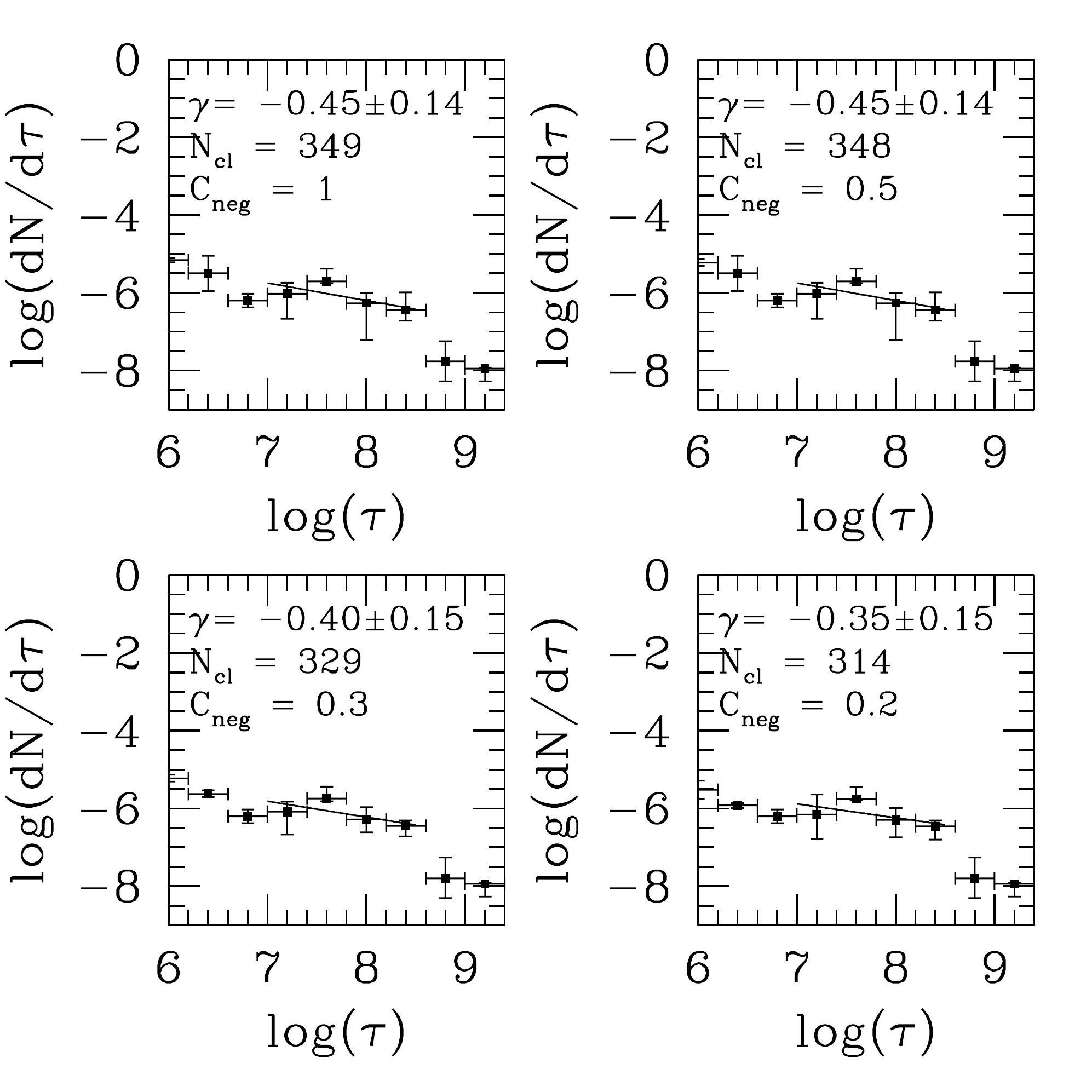}
\caption{The age distributions for Class 1 and 2 clusters in the C field using our \starcnet catalog. From the top left to the bottom right panel we increase the threshold for objects to be included as clusters in our final catalog from 0-80 \% confidence. The resulting distribution slope $\gamma$ appears to slightly flatten as the thresholds increase and the number of clusters in the overall sample decreases, but demonstrates that the age distribution inferred from our \starcnet catalogs is insensitive to the choice of $C_{neg}$ adopted in our analysis.}
\end{figure*}

\section{Summary}

We present {\it Hubble Space Telescope} WFC3/UVIS (F275W, F336W) and ACS/WFC optical (F435W, F555W, and F814W) observations of the nearby grand-design spiral galaxy M101 as part of the Legacy Extragalactic UV Survey. Using Source Extractor we build an initial catalog of 39,705 objects with a S/N of $\geq 10$ for 10 pixels in at least 2 of our photometric bands. 2,351 of these sources with $m_{V} < 23.14$ were visually inspected and assigned a classification of 1 (symmetric), 2 (asymmetric), 3 (multi-peak; associations), or 4 (single stars, galaxies, etc.). We also utilize the convolutional neural network (CNN) \starcnet to provide classifications on the same scale for the entire catalog of objects. Class 1 and 2 sources are retained as clusters in our final catalogs. Finally, we impose a requirement of 4 or more photometric band detections and $\sigma_{U} \leq 0.1$ mag for each source in order to ensure we can extract robust ($\sigma_{age,mass} \sim 0.2$ dex) estimates for cluster ages, masses and extinctions. With both our human-classified ($N_{c} = 965$) and \starcnet-classified ($N_{c} = 2,270$) cluster catalogs we have produced the most complete census to date of compact star clusters in M101. The following conclusions are reached:

\noindent
(1) For the 2,351 sources with both a visual- and ML-classification \starcnet is able to reproduce the human classifications at high levels of accuracy ($\sim 80-90\%$) for binary classification (cluster vs non-cluster), which is equivalent to the level of agreement between human classifiers in LEGUS. In particular \starcnet appears to be able to recover several faint Class 1 objects that are mis-classified as Class 4 in the human-generated catalog. By comparing the magnitude distributions for each catalog using both classification methods we find that indeed the vast majority of the Class 1 and 2 clusters added in the \starcnet catalog relative to the human-classified catalog is a population of faint sources which were mostly missed by the limitations of visual inspection.

\noindent
(2) The derived cluster age distribution implies a disruption rate of $dN/d\tau \propto \tau^{-0.45 \pm 0.14}$ over $10^{7} < \tau < 10^{8.5}$ for cluster masses $\geq 10^{3.55} M_{\odot}$ for the central region of M101 and $dN/d\tau \propto \tau^{-0.02 \pm 0.15}$ for the northwest region of the galaxy. This is consistent with observations of other nearby spiral galaxies which show enhancements in the cluster disruption rate as a function of decreasing galactocentric radius. Finally, we find that splitting our age distributions by mass ranges results in steeper age distribution slopes toward lower masses. These results both provide evidence in favor of environmentally-dependent cluster disruption in the central, southeast, and northwest regions of M101. They also hint at the possibility of mass-dependent disruption, although our data do not provide a definite answer to this issue given the completeness limits of our cluster sample, and the relatively limited coverage of the total galaxy stellar populations (Figure 1).

\noindent
(3) The derived cluster masses imply CMFs for M101 which are all well-described as $dN/dM \propto M^{-2}$ across our three regions, two age intervals, and both the human and ML-classified catalogs. This is consistent with the canonical -2 power law found for nearly all normal star-forming galaxies in the local Universe.

\noindent
(4) The age distributions inferred from the human-classified cluster distributions suffer from a bias in the objects selected for inspection. Further, we show that the \starcnet Class 1 and 2 classifications are robust against a range of confidence thresholds for each catalog. Therefore the use of these catalogs does not bias our results in any significant way, and represents the best path forward for creating fast ML-generated catalogs of clusters in large numbers of nearby galaxies which now have deep multi-band HST archival imaging and will soon be observed with the James Webb Space Telescope. The final cluster catalogs for M101 are available on the LEGUS website (https://legus.stsci.edu).

\acknowledgements

S.T.L. acknowledges partial support from the NASA grant HST GO--15330. Support for program \# 15330 was provided by NASA through a grant from the Space Telescope Science Institute. K.G. is supported by the Australian Research Council through the Discovery Early Career Researcher Award (DECRA) Fellowship
DE220100766 funded by the Australian Government. K.G. is supported by the Australian Research Council Centre of Excellence for All Sky Astrophysics in 3 Dimensions (ASTRO~3D), through project number CE170100013. M.M. acknowledges the support of the Swedish Research Council, Vetenskapsradet (internationell postdok grant 2019-00502). Based on observations made with the NASA/ESA Hubble Space Telescope, obtained  at the Space Telescope Science Institute, which is operated by the  Association of Universities for Research in Astronomy, Inc., under NASA contract NAS 5--26555. These observations are associated with program \# 13364. Based also on archival data from the following facilities: the NASA/ESA Hubble Space Telescope, and obtained from the Hubble Legacy Archive, which is a collaboration between the Space Telescope Science Institute (STScI), the Space Telescope European Coordinating Facility (ST-ECF/ESA) and the Canadian Astronomy Data Centre (CADC/NRC/CSA). Finally, This research has made use of the NASA/IPAC Extragalactic Database (NED) which is operated by the Jet Propulsion Laboratory, California Institute of Technology, under contract with the National Aeronautics and Space Administration.

\bibliography{master_ref}

\appendix

\begin{figure*}
  \centering
  \includegraphics[scale=0.4]{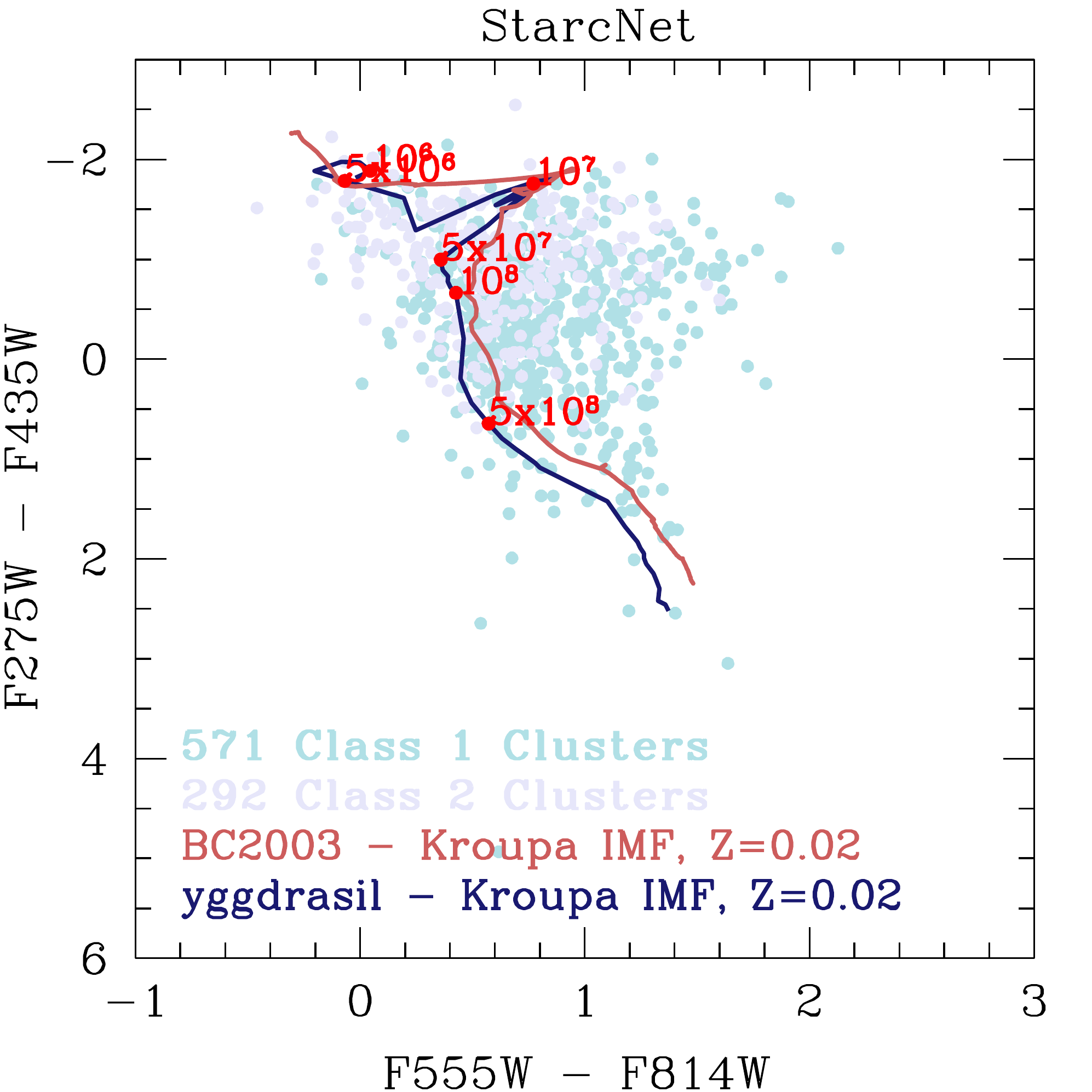}
  \includegraphics[scale=0.4]{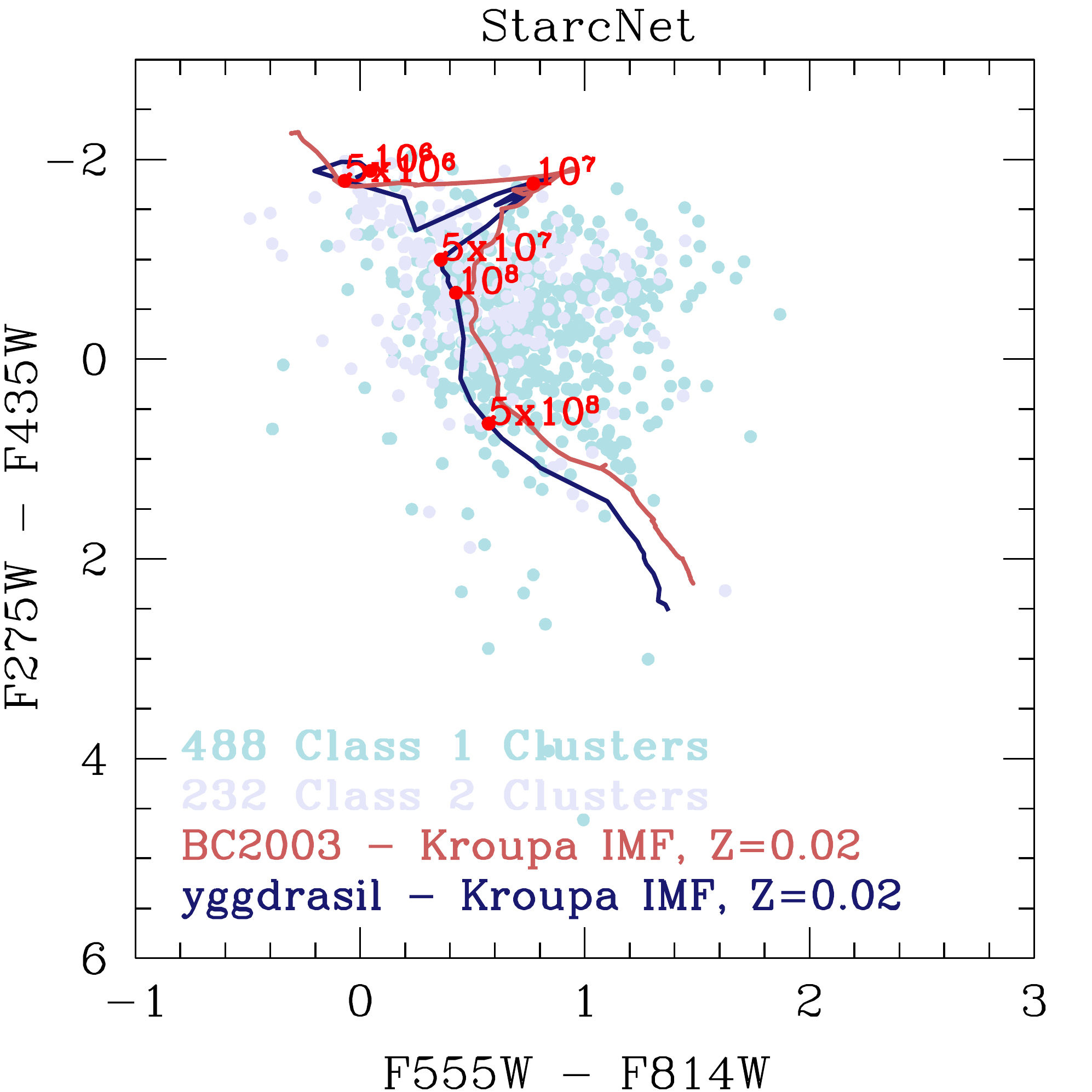}
  \includegraphics[scale=0.4]{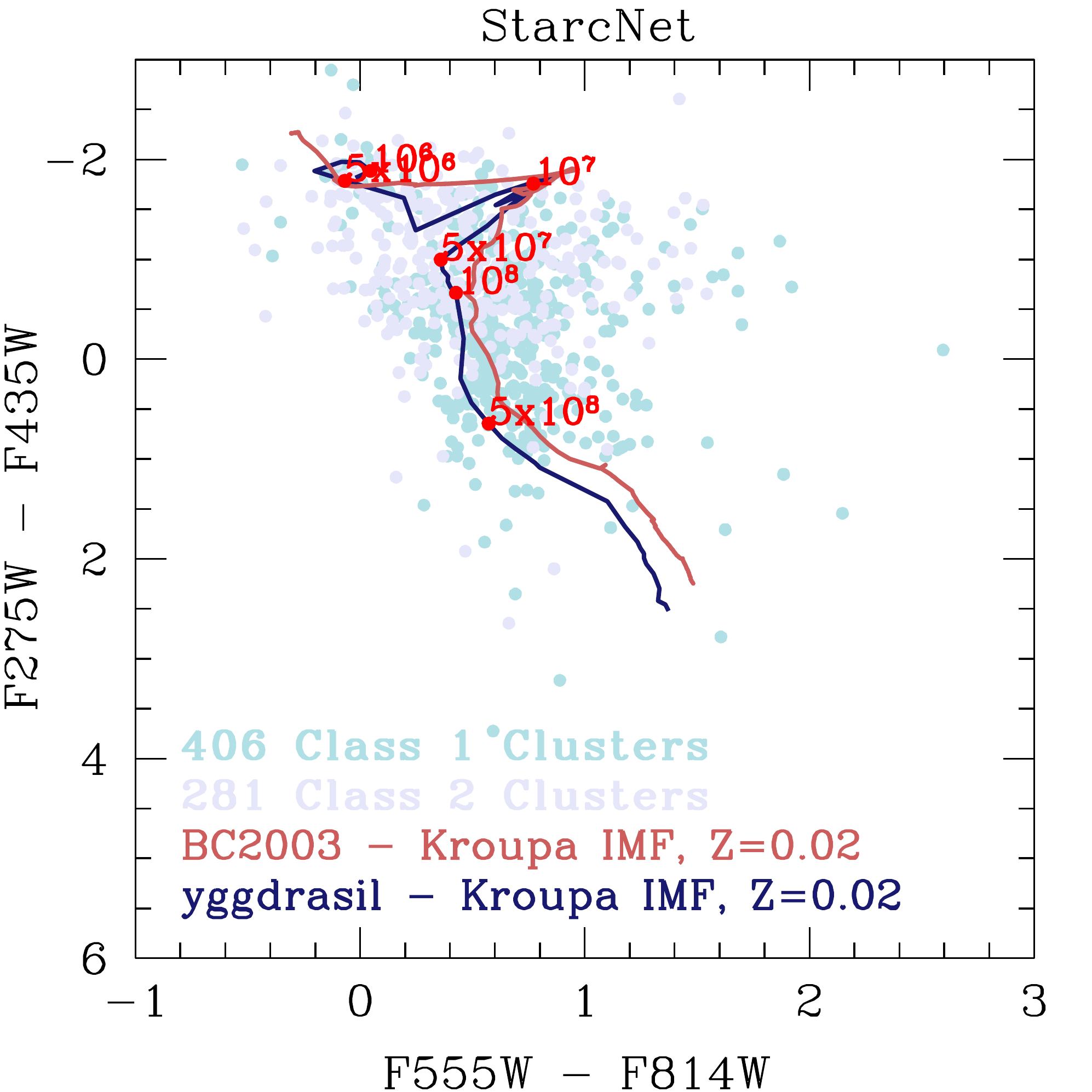}
  \caption{Same as Figure 3 for the central (top left), southeast (top right), and northwest (bottom) regions respectively. As discussed in \S 2.2 The differences in the observed NUV-B vs. V-I color distributions can be attributed to the differences in the cluster disruption rates measured in the C and NW regions \citep{nb11}.}
\end{figure*}

\begin{figure*}
  \centering
  \includegraphics[scale=0.4]{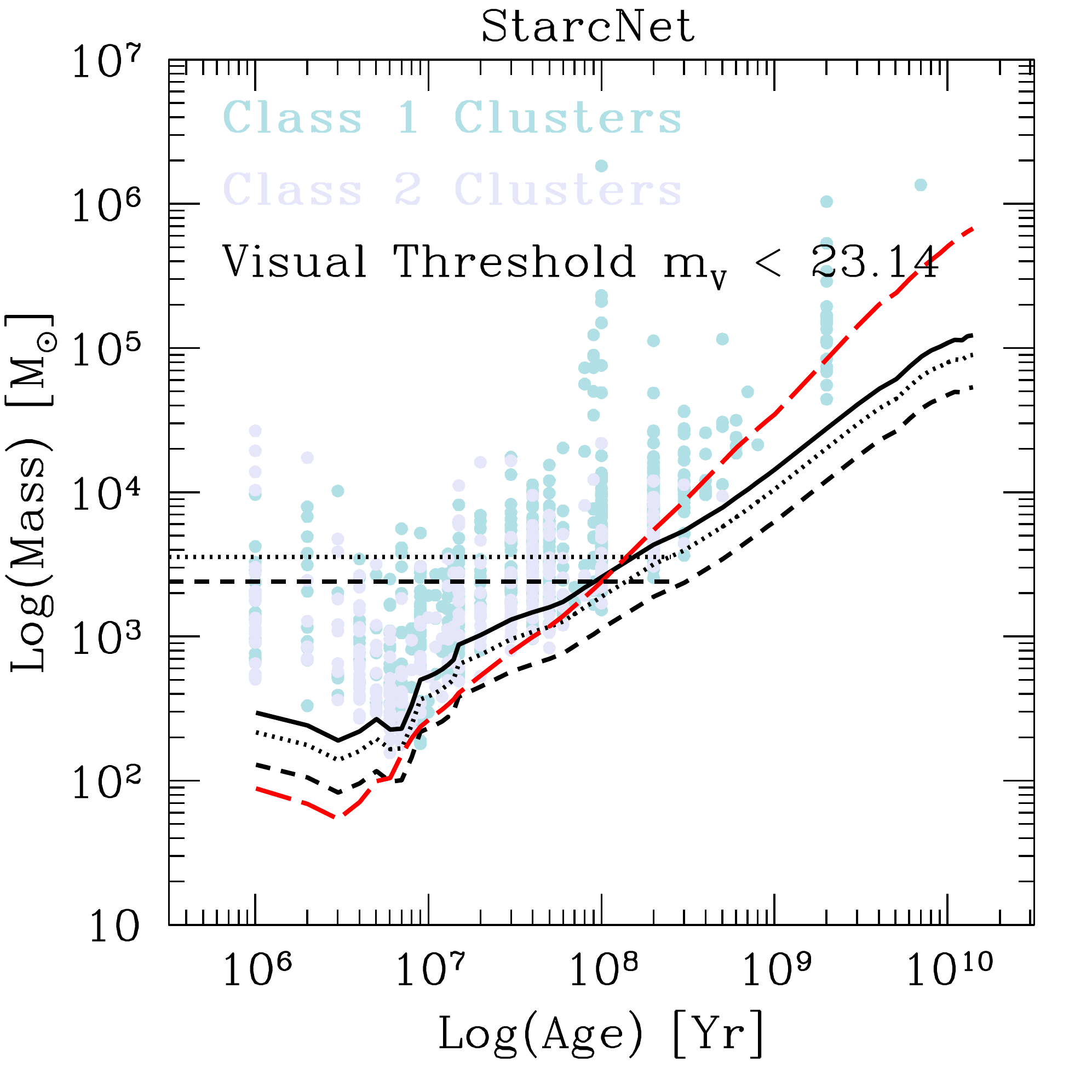}
  \includegraphics[scale=0.4]{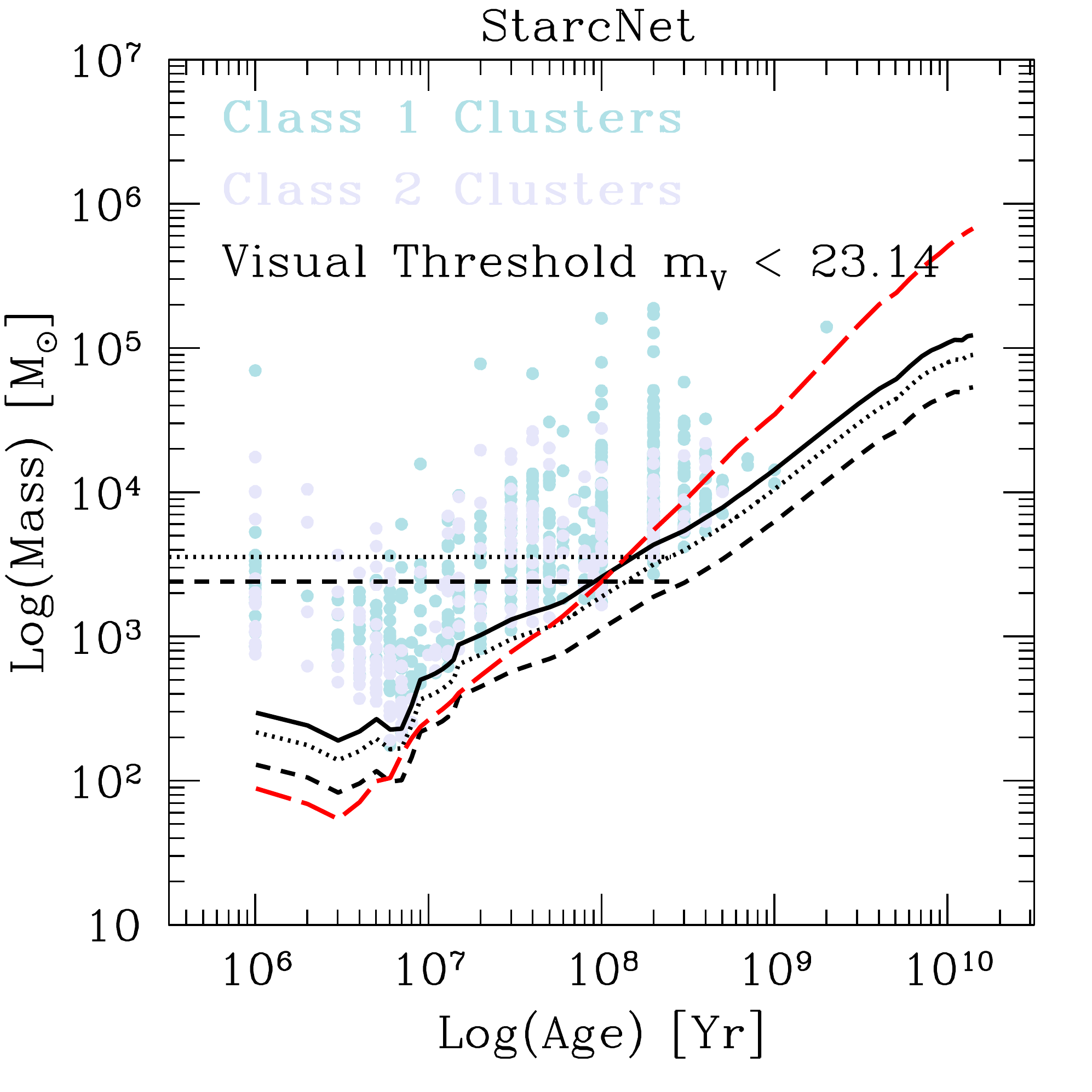}
  \includegraphics[scale=0.4]{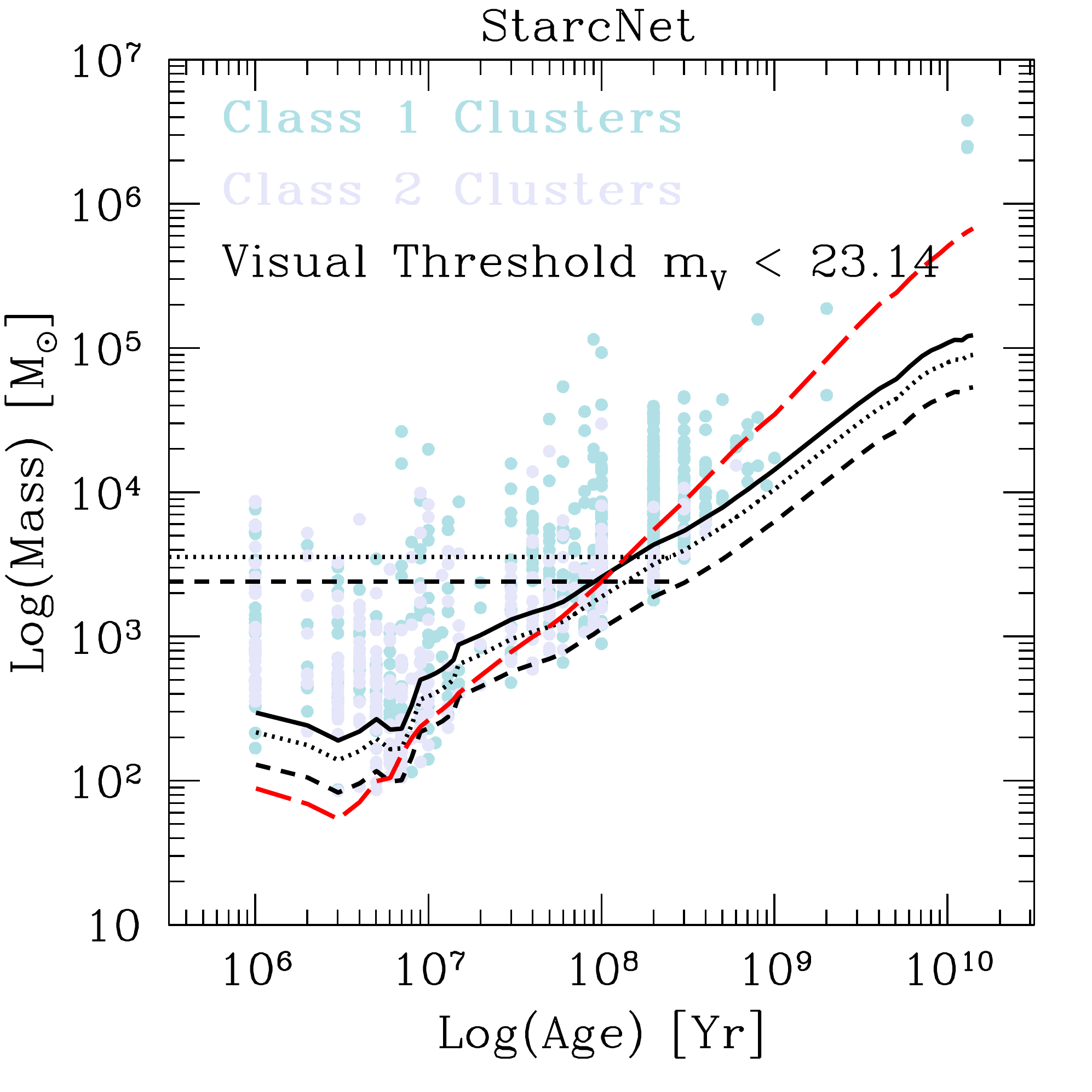}
  \caption{Same as Figure 5 for the central (top left), southeast (top right), and northwest (bottom) regions respectively. As discussed in \S 2.2 the northwest region includes fainter clusters than the other two regions because it is less affected by crowding.}
\end{figure*}

\begin{figure*}
  \centering
  \includegraphics[scale=0.4]{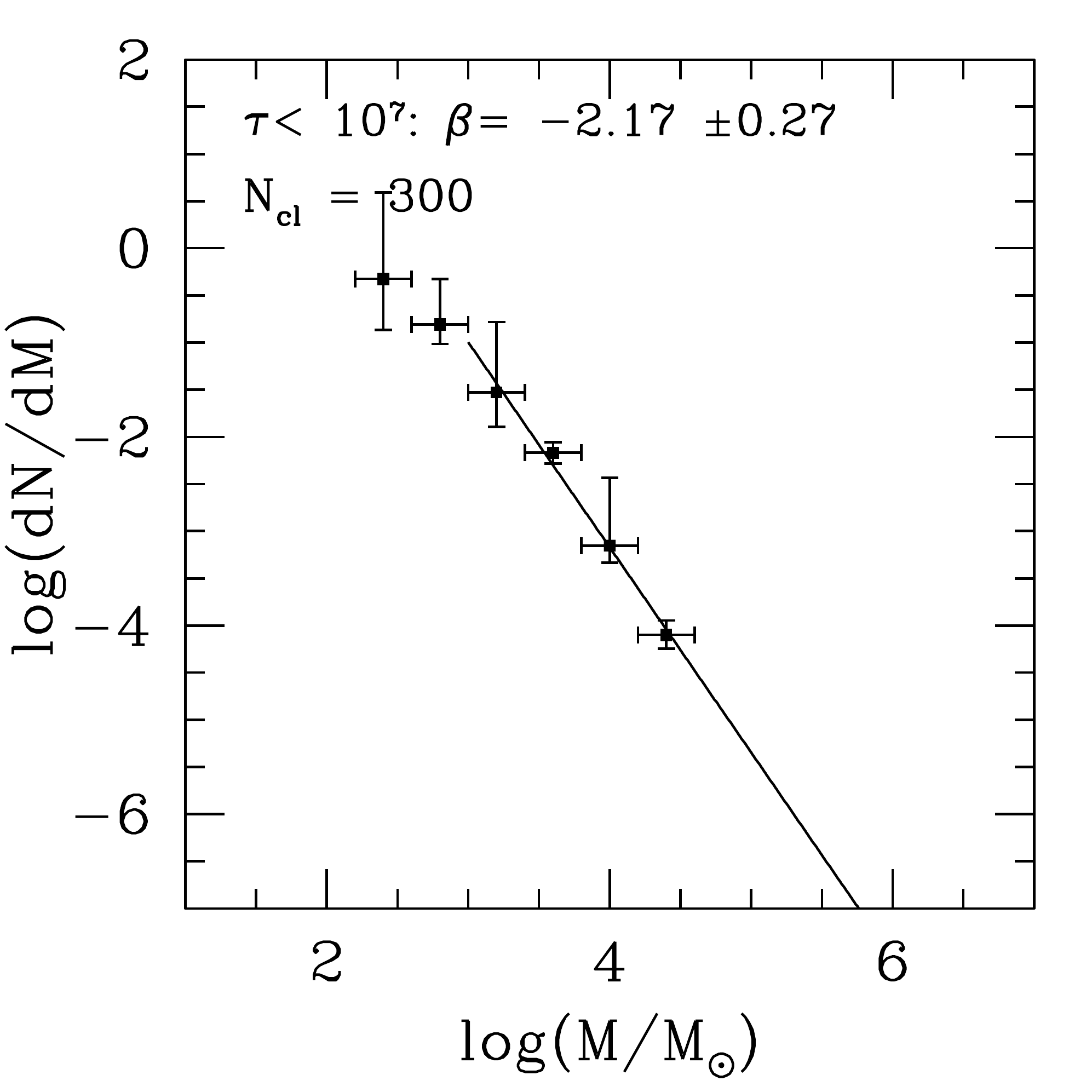}
  \includegraphics[scale=0.4]{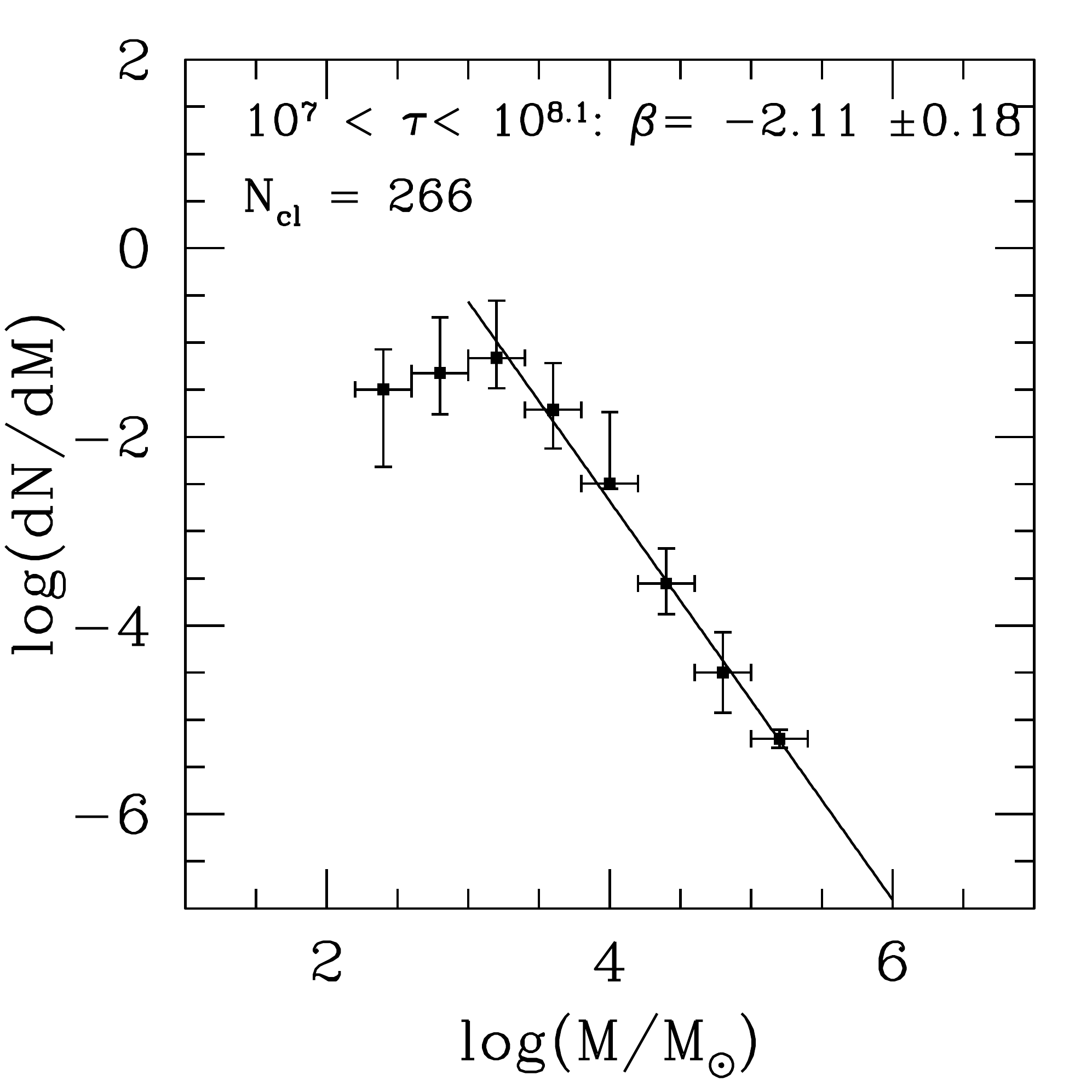}
  \caption{Same as Figure 9 for the young (left) and intermediate-age (right) clusters in the southeast region.}
\end{figure*}

\begin{figure*}
  \centering
  \includegraphics[scale=0.4]{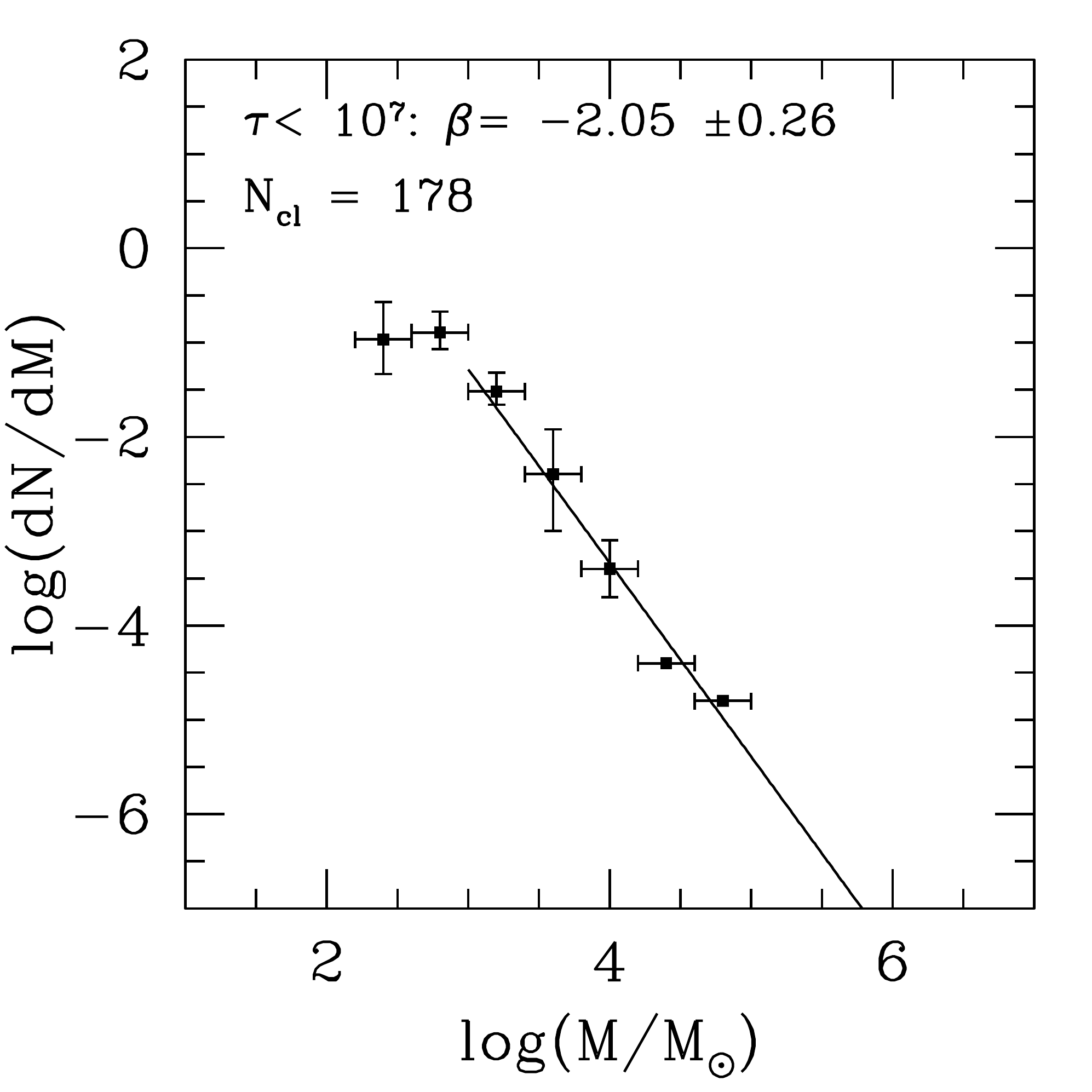}
  \includegraphics[scale=0.4]{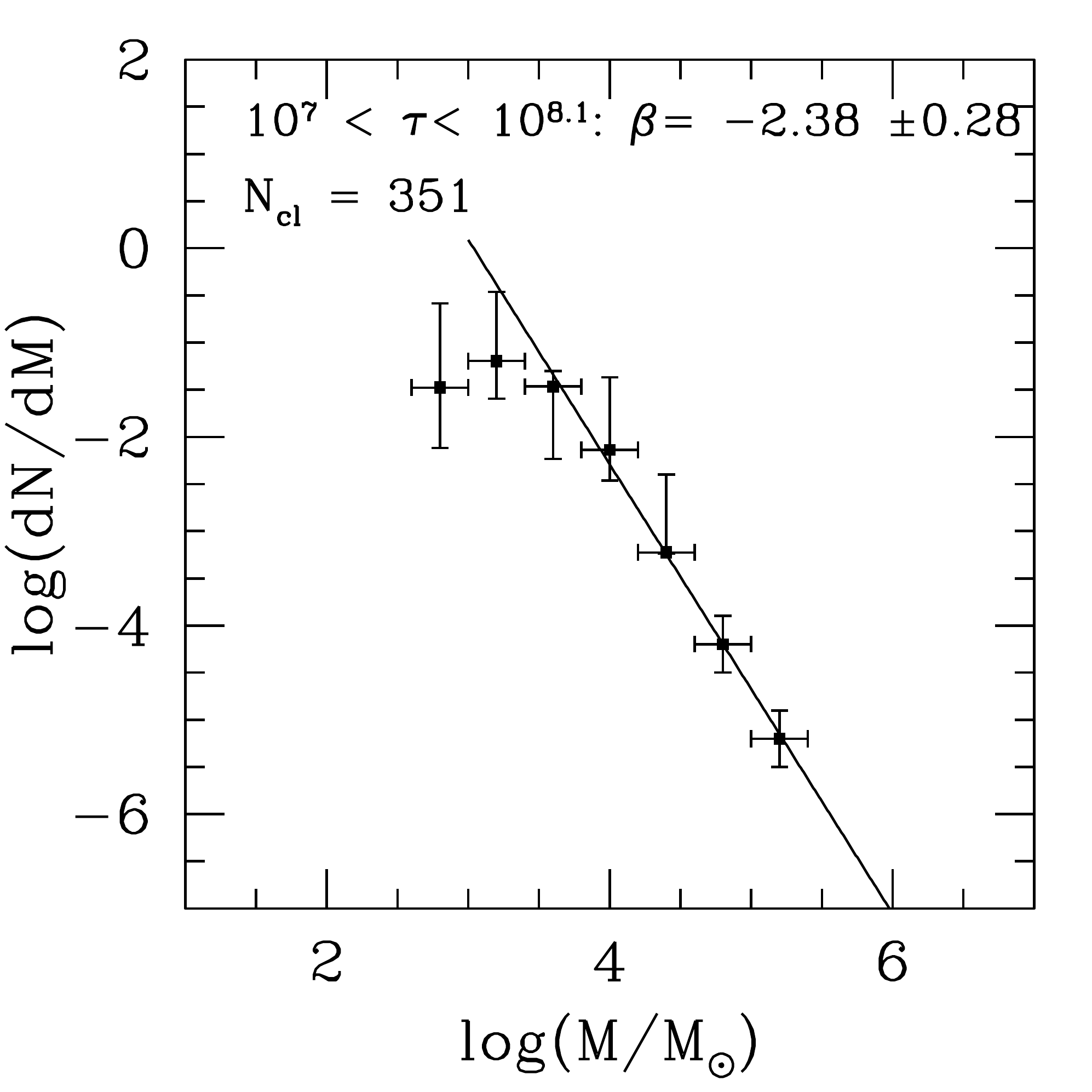}
  \caption{Same as Figure 9 for the young (left) and intermediate-age (right) clusters in the northwest region.}
\end{figure*}

\begin{figure*}
\centering
     \includegraphics[scale=0.65]{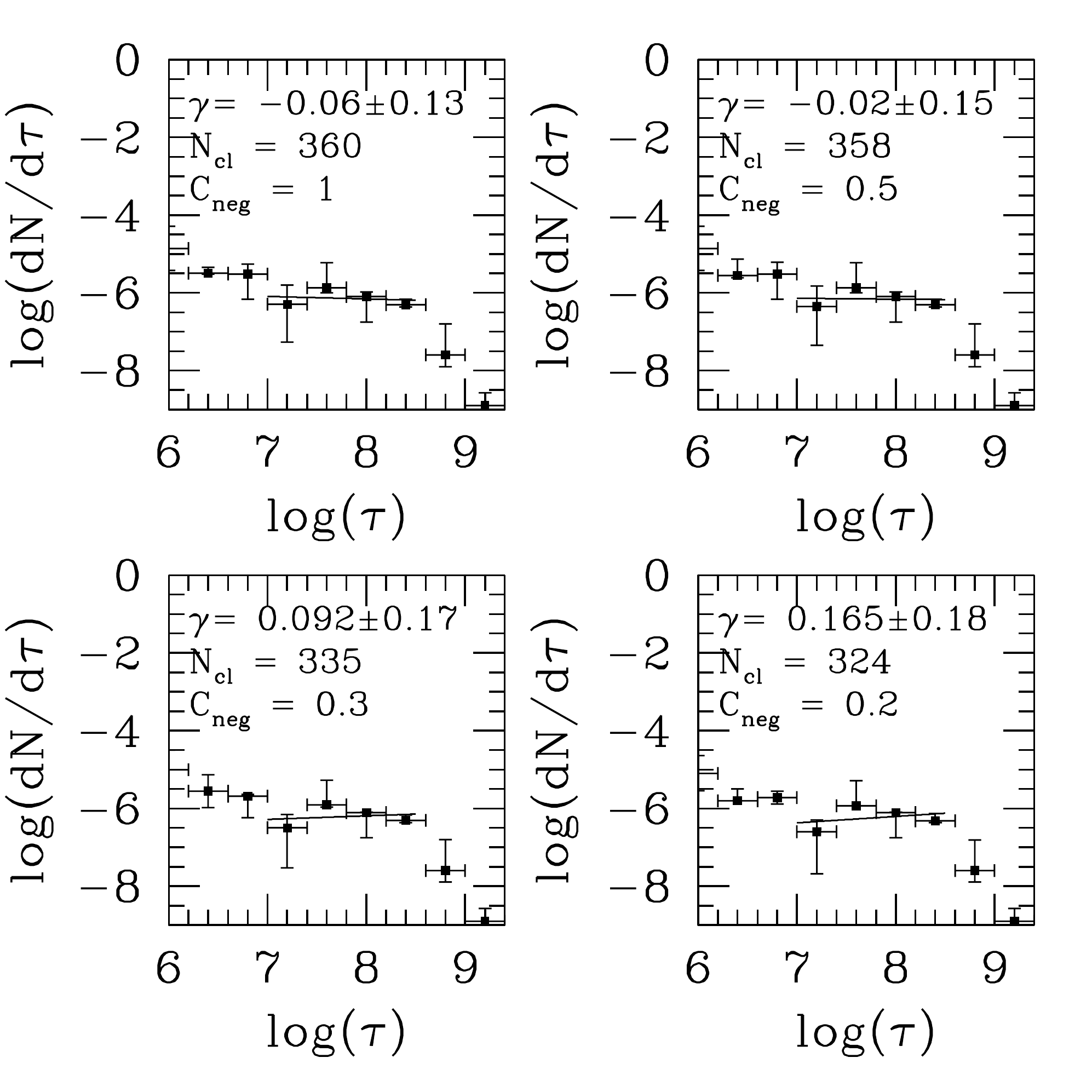}
\caption{Same as Figure 9 for the NW catalog.}
\end{figure*}

\begin{figure*}
\centering
     \includegraphics[scale=0.65]{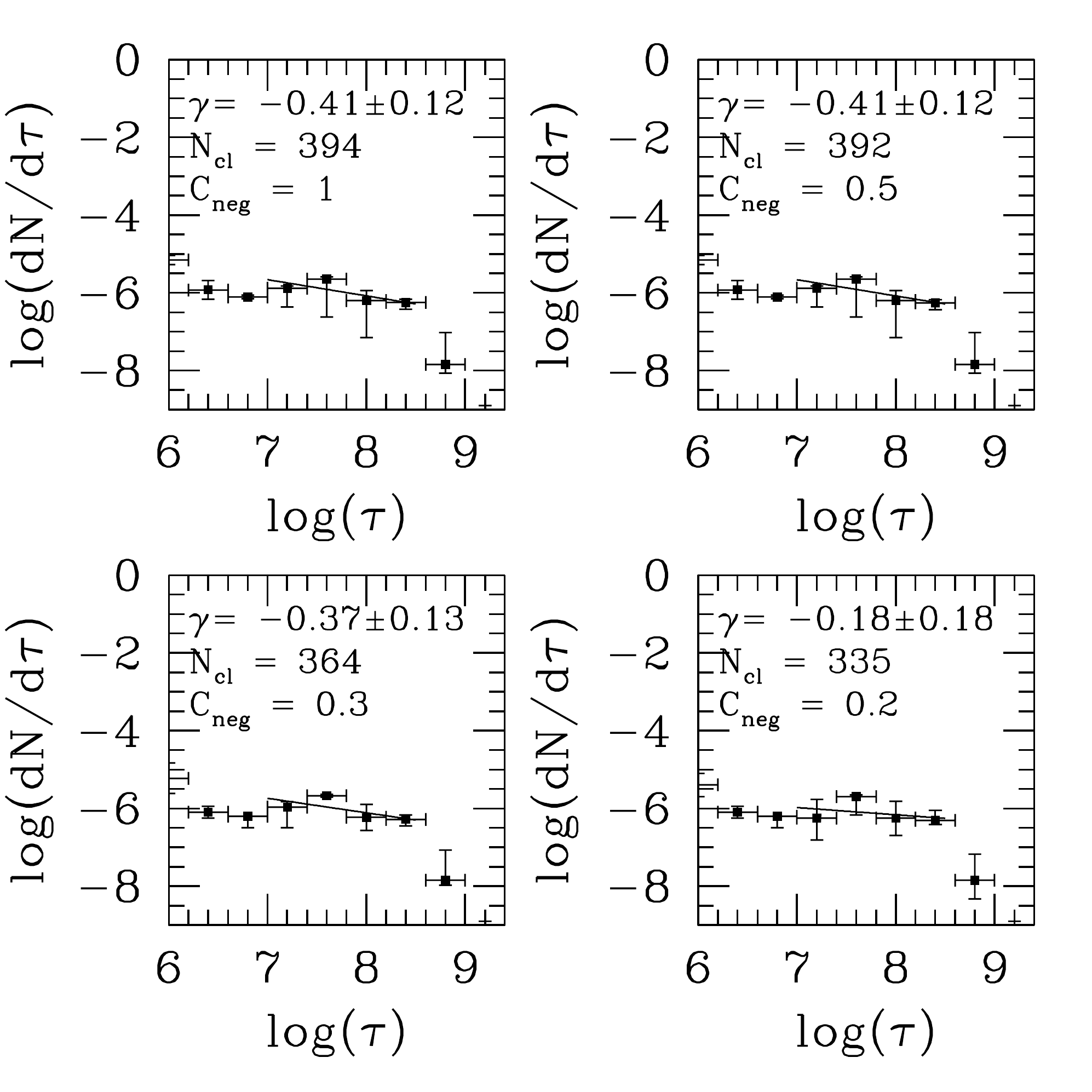}
\caption{Same as Figure 9 for the SE catalog.}
\end{figure*}

\end{document}

%% file: tbl-1.tex
\begin{deluxetable*}{l|ccccccccc}
\center
\tablecaption{Cluster Detection Statistics and Overall Properties \label{tbl-1}}
\tabletypesize{\footnotesize}
\tablewidth{0pt}
\tablehead{
\colhead{Catalog}  & \colhead{$N_{det}$} & \colhead{$N_{c},ml$} & \colhead{$N_{c},vis$} & \colhead{$\tau_{c}$} & \colhead{$\sigma_{\tau_{c}}$} & \colhead{$M_{c}$} & \colhead{$\sigma_{M_{c}}$} &  \colhead{$E (B-V)$} & \colhead{$\sigma_{E (B-V)}$}}
\startdata
Central & 1825 & 863 & 354 & 7.60 & 0.52 & 3.43 & 0.36 & 0.17 & 0.12 \\
Southeast & 1493 & 720 & 385 & 7.61 & 0.63 & 3.60 & 0.33 & 0.17 & 0.14 \\
Northwest & 1417 & 687 & 226 & 7.60 & 0.61 & 3.34 & 0.49 & 0.16 & 0.14 \\
Total & 4725 & 2270 & 965 & 7.61 & 0.58 & 3.46 & 0.40 & 0.17 & 0.13 \\
\enddata
\tablecomments{$N_{det}$ is the number of cluster candidates in extracted in each catalog detected in four bands. $N_{c},ml$ and $N_{c},vis$ are the number of confirmed Class 1+2 clusters using the \starcnet ML algorithm and visual inspection respectively. Values for the median cluster age ($\tau$), mass, and extinction as well as their median absolute deviation (MAD) in the \starcnet-generated catalogs are given in log($\tau$yr), log($M/M_{\odot}$), and magnitudes respectively.}
\end{deluxetable*}

%% file: tbl-2.tex
\begin{deluxetable*}{l|cccc}
\center
\tablecaption{Mass-Dependent Age Distribution Fits\label{tbl-2}}
\tabletypesize{\footnotesize}
\tablewidth{0pt}
\tablehead{
\colhead{Sample}  & \colhead{$N_{c}: 10^{3.4} - 10^{3.6} M_{\odot}$} & \colhead{$N_{c}: 10^{3.6} - 10^{3.8} M_{\odot}$} & \colhead{$N_{c}: 10^{3.8} - 10^{4.0} M_{\odot}$} & \colhead{$N_{c}: 10^{4.0} - 10^{4.4} M_{\odot}$}}
\startdata
Central & 130 & 122 & 74 & 76  \\ 
Southeast & 93 & 112 & 112 & 107   \\ 
Northwest & 88 & 74 & 72 & 78   \\ 
\enddata
\tablecomments{$N_{c}$ is the number of Class 1 + 2 clusters with masses between the stated limits in each column for each of the three regions in M101.}
\end{deluxetable*}

%% file: tbl-3.tex
\begin{deluxetable}{l|ccc}
\center
\tablecaption{Results From MSPECFIT for the StarcNet Catalogs\label{tbl-3}}
\tabletypesize{\footnotesize}
\tablewidth{0pt}
\tablehead{
\colhead{Sample}  & \colhead{$N_{c}$} & \colhead{$\beta$} & \colhead{$\sigma_{\beta}$}}
\startdata
Central-Young & 245 & -2.27 & 0.19  \\ 
Central-Int & 451 & -2.06 & 0.24  \\ 
Southeast-Young & 300 & -2.17 & 0.27  \\ 
Southeast-Int & 266 & -2.11 & 0.18  \\ 
Northwest-Young & 178 & -2.05 & 0.26  \\ 
Northwest-Int & 351 & -2.38 &  0.28 \\ 
\enddata
\tablecomments{$N_{c}$ is the number of Class 1 + 2 clusters with masses above the completeness limits and within the stated age ranges. Young and intermediate clusters are defined as having ages $\tau < 10^{7}$yr and $10^{7}$yr$ < \tau < 10^{8}$yr. $\beta$ is the slope of the PL fit the CMF using MSPECFIT.}
\end{deluxetable}